\documentclass[apj,numberedappendix]{emulateapj}
\usepackage{epstopdf}
\usepackage{wrapfig}
\usepackage{natbib}
\bibpunct{(}{)}{;}{a}{}{,}

\usepackage{color}
\definecolor{darkgreen}{RGB}{0,142,128}
\definecolor{darkblue}{RGB}{0,100,170}

\usepackage{hyperref}
\hypersetup{colorlinks,citecolor=blue,linkcolor=blue}

\begin{document}
\title{Age dependence of wind properties for solar type stars: a 3D study}

\author{Victor R\'eville$^1$,	
	Colin P. Folsom$^{2,3}$,
	Antoine Strugarek$^{4,1}$,
        Allan Sacha Brun$^1$      
}

\affil{$^1$Laboratoire AIM, DSM/IRFU/SAp, CEA Saclay, 91191 Gif-sur-Yvette Cedex, France; victor.reville@cea.fr, sacha.brun@cea.fr\\
$^2$ IPAG, Universit\'e Joseph Fourier, B.P.53 F-38041 Grenoble Cedex 9-France, colin.folsom@obs.ujf-grenoble.fr\\
$^3$ IRAP, CNRS - Universit\'e de Toulouse, 14 avenue Edouard Belin 31400 Toulouse \\
$^4$ D\'epartement de physique, Universit\'e de Montr\'eal, C.P. 6128 Succ. Centre-Ville, Montr\'eal, QC H3C-3J7, Canada; strugarek@astro.umontreal.ca\\
}

\begin{abstract}

Young and rapidly rotating stars are known for intense, dynamo generated magnetic fields. Spectropolarimetric observations of those stars in precisely aged clusters are key input for gyrochronology and magnetochronology. We use ZDI maps of several young K-type stars of similar mass and radius but with various ages and rotational periods, to perform 3D numerical MHD simulations of their coronae and follow the evolution of their magnetic properties with age. Those simulations yield the coronal structure as well as the instant torque exerted by the magnetized, rotating wind on the star. As stars get older, we find that the angular momentum loss decreases with $\Omega_{\star}^3$, which is the reason for the convergence on the Skumanich law. For the youngest stars of our sample, the angular momentum loss show signs of saturation around $8 \Omega_{\odot}$, which is a common value used in spin evolution models for K-type stars. We compare these results to semi-analytical models and existing braking laws. We observe a complex wind speed distribution for the youngest stars with slow, intermediate and fast wind components, which are the result of the interaction with intense and non axisymmetric magnetic fields. Consequently, in our simulations, the stellar wind structure in the equatorial plane of young stars varies significantly from a solar configuration, delivering insight about the past of the solar system interplanetary medium.
\end{abstract}

%\keywords{Stuff, Stars: stuff}

\section{Introduction} 
\label{intro}

Among all the stellar properties, the characteristics of solar-like stars' winds are probably the most difficult to probe. Emissions throughout the electromagnetic spectrum unveil some of the properties of the photosphere and the coronae of stars, and internal structures can be probed with asteroseismology. Winds, on the contrary, produce very few detectable signals, although they are likely to exist in all stars possessing a hot corona, as supersonic outflows are the only stable way to balance coronal pressure with the near zero interstellar medium pressure \citep{Parker1958,Velli1994}. \citet{LinskyWood1996} have shown that absorption by neutral hydrogen at the astropause could be detected in Ly{$\alpha$} spectra around astrospheres of nearby solar-like stars, unraveling properties of the stellar wind shocking against the interstellar medium. A growing sample of solar-type stars with positive detection for stellar winds led to a relationship between X-ray fluxes originating from coronal loops and mass loss rates \citep{Wood2002}.  The ``strength" of stellar winds, the mass loss rate $\dot{M}$, has consequently been related to the magnetic activity of the star. \citet{Wood2005a} have obtained the relation: $\dot{M} \propto F_X^{1.34 \pm 0.18}$, for $F_X \leq 10^6$ ergs cm$^{-2}$ s$^{-1}$, where $F_X$ is the X-ray flux. Beyond this value, weaker mass loss rates are observed, suggesting a saturation effect, that is below the usual $F_X$ saturation value \citep{Randich2000,Pizzolato2003,Gudel2004}.

In parallel, the development of Zeeman Doppler Imaging (ZDI) \citep{Semel1989,DonatiBrown1997,PiskunovKochukhov2002} has opened the study of surface magnetic fields for cool stars. Large scale magnetic fields have been detected in the whole mass range that is thought to correspond to the existence of a convective envelope ($0.1 M_{\odot} - 1.4 M_{\odot}$). The study of the geometrical and topological properties of the field in the context of stellar evolution is still in progress \citep{DonatiLandstreet2009,See2015} and raises theoretical questions about their generation through dynamo processes in convective envelopes \citep[see][and references therein]{Brun2015}. Nonetheless, the magnetic field amplitude of these stars has been shown to be a growing function of the rotation rate \citep{Noyes1984,BrandenburgSaar2000,Vidotto2014b}. This is necessary to explain the rotational braking of cool main sequence dwarfs, as evolutionary models need the wind to carry angular momentum at a rate proportional to $\Omega_{\star}^3$ \citep{Kawaler1988,Bouvier1997,Matt2015} all along the main sequence. However, recent studies suggest that the wind braking could stop or strongly decay for evolved stars, around a solar Rossby number $Ro \approx 2.5$ \citep{vanSaders2016}, perhaps because of a change in magnetic topology \citep{Reville2015a,Garraffo2015a}. Hence wind, magnetism, and rotation are likely to evolve coherently through the whole life of solar-like stars.

After \citet{Schatzman1962} understood that a magnetized outflow would carry away stellar angular momentum, \citet{WeberDavis1967} demonstrated that this loss is proportional to the Alfv\'en radius squared. Several studies have followed to try to estimate the Alfv\'en radius from stellar parameters and thus give scaling laws for the angular momentum loss \citep{Mestel1968,Kawaler1988}. The latest braking laws have been successfully introduced in stellar evolution models \citep{Matt2012,GalletBouvier2013}, and we recently demonstrated that the magnetic topology could be included in those formulations through a simple scalar parameter, the open magnetic flux \citep{Reville2015a,Reville2015b}.

Most studies \citep[see, e.g.,][]{Matt2012,Reville2015a} have been made in two dimensions with axisymmetric configurations \citep[see][for a 3D study of angular momentum loss with idealized magnetic field topologies]{Cohen2014,Garraffo2015a}, and were not able to capture the structure of complex magnetic fields observed by ZDI. 3D MHD simulations are now taking into account this complexity \citep{Cohen2011,Vidotto2014a,doNascimento2016,AlvaradoGomez2016a,AlvaradoGomez2016b} to derive a self-consistent coronal structure. The complex structure of the corona is needed to study the interaction between stars and close-in planets, which has been shown to be very sensitive to 3D effects \citep{Strugarek2015}. Yet, to our knowledge, the influence of realistic magnetic fields on the long time variation of the wind properties has not been studied.

This work proposes to include observed, realistic magnetic fields in the context of stellar evolution. We used spectropolarimetric observations of the surface fields of solar-like stars to constrain 3D MHD simulations of stellar winds. The stars of our sample share similar properties except their rotational periods and their ages, which range from  25 Myr to 4.5 Gyr. We developed a coherent framework to take into account the evolution of the coronal properties with time, inspired by X-ray flux observations, spin evolution models and theoretical, ab initio models \citep[see][]{HolzwarthJardine2007,CranmerSaar2011,Suzuki2013b}. We confirm that the evolution with age of global properties of the wind, such as the mass and angular momentum loss, follows simple prescriptions in agreement with the spin evolution models. These prescriptions can be recovered by the semi-analytical model we developed in \citet{Reville2015b}, except for the saturation of angular momentum that appears only in our simulations. Also, the three dimensional structure of the young stars' winds shows interesting features, such as a trimodal speed distribution, that we explain through various interactions with the intense magnetic field. We show that superradial expansion is a key factor to explain the fastest wind components of young stars' wind. We also observe regions of fast wind encountering slower streams in the equatorial plane, the so-called Corotating Interactions Regions (CIRs), that could be more common in the wind of young stars.

This paper is organized as follows: the ingredients of our model are described throughout Section \ref{sec:model}. In Subsection \ref{subsec:zdi}, we describe the observations used to constrain the surface magnetic fields of our simulations. Subsection \ref{subsec:evol} describes our prescriptions for the coronal properties and Subsection \ref{subsec:num} our numerical setup. The results are presented in two parts, Section \ref{sec:glob} where we look at global properties such as the angular momentum and mass loss over time, and Section \ref{sec:vel} where the tridimensional structure of the wind is detailed, with a special focus on the velocity distribution. We summarize and reflect upon our findings in Section \ref{sec:ccl}.

\section{Model Ingredients and Description}
\label{sec:model}

\subsection{Observational data: ZDI maps}
\label{subsec:zdi}

\begin{deluxetable*}{l|c|c|c|c|c|c|c|c}
  \tablecaption{Stellar parameters of the Young Solar Analogs}
  \tablecolumns{9}
  \tabletypesize{\scriptsize}
  \tablehead{
    \colhead{Name} &
    \colhead{Age (Myr)} &
    \colhead{Period (days)} &
    \colhead{Mass ($M_{\odot}$)} &
    \colhead{Radius ($R_{\odot}$)} &
    \colhead{$T_{\mathrm{eff}}$ (K)} &
    \colhead{$\langle B_r \rangle$ (G)} &
    \colhead{\% dipole} &
    \colhead{\% axis.}
  }
  \startdata

  BD- 16351 & $42\pm 6$ & $3.3$ & $0.9$ & $0.9$ & $5243$ & $33.0$ & $60$ & $5$\\
  TYC 5164-567-1 & $120\pm 10$ & $4.7$ & $0.9$ & $0.9$ & $5130$ & $48.8$ & $78$ & $78$\\
  HII 296 & $125\pm 8$ & $2.6$ & $0.9$ & $0.9$ & $5322$ & $52.0$ & $57$ & $50$ \\
  DX Leo & $257\pm 47$ & $5.4$ & $0.9$ & $0.9$ & $5354$ & $21.3$ & $69$ & $1$ \\
  AV 2177 & $584 \pm 10$ & $8.4$ & $0.9$ & $0.9$ & $5316$ & $5.4$ & $57$ & $4$\\
  Sun 1996 & $4570$ & $28$ & $1.0$ & $1.0$ & $5778$ & $1.1$ & $35$ & $75$ \\
  \enddata
  \tablecomments{See \citet{Folsom2016} and references therein. The study of AV 2177 belongs to the second part of the aforementioned study.}
  \label{table1}
\end{deluxetable*}

\begin{figure*}
\begin{tabular}{rr}
\includegraphics[scale=0.5]{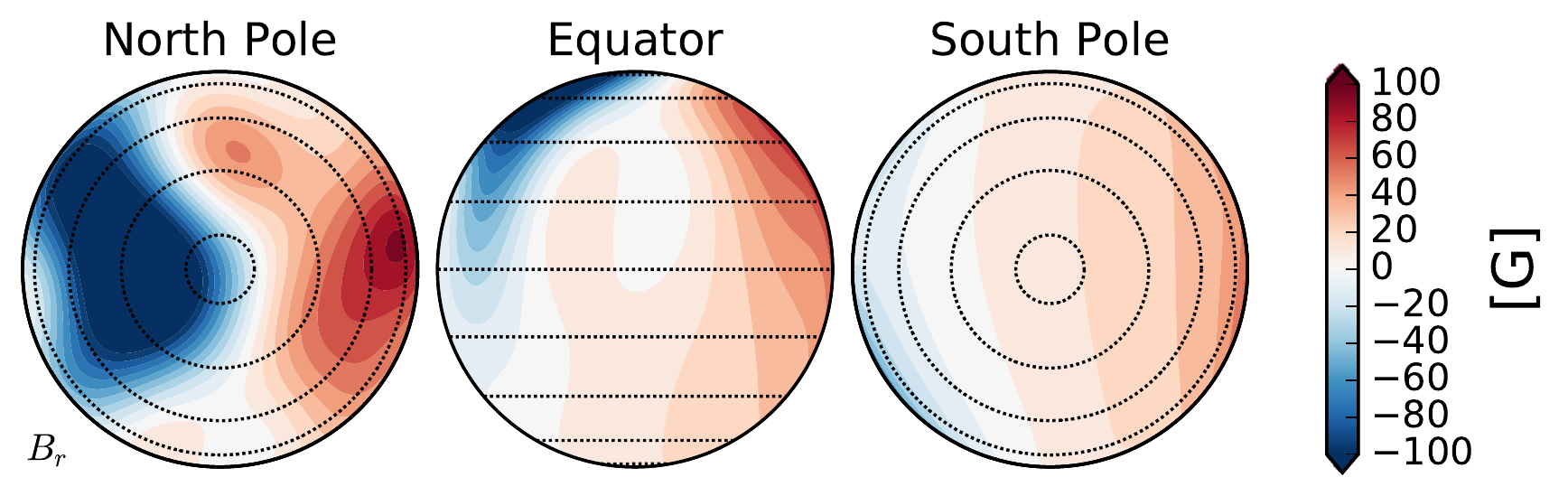} & \includegraphics[scale=0.5]{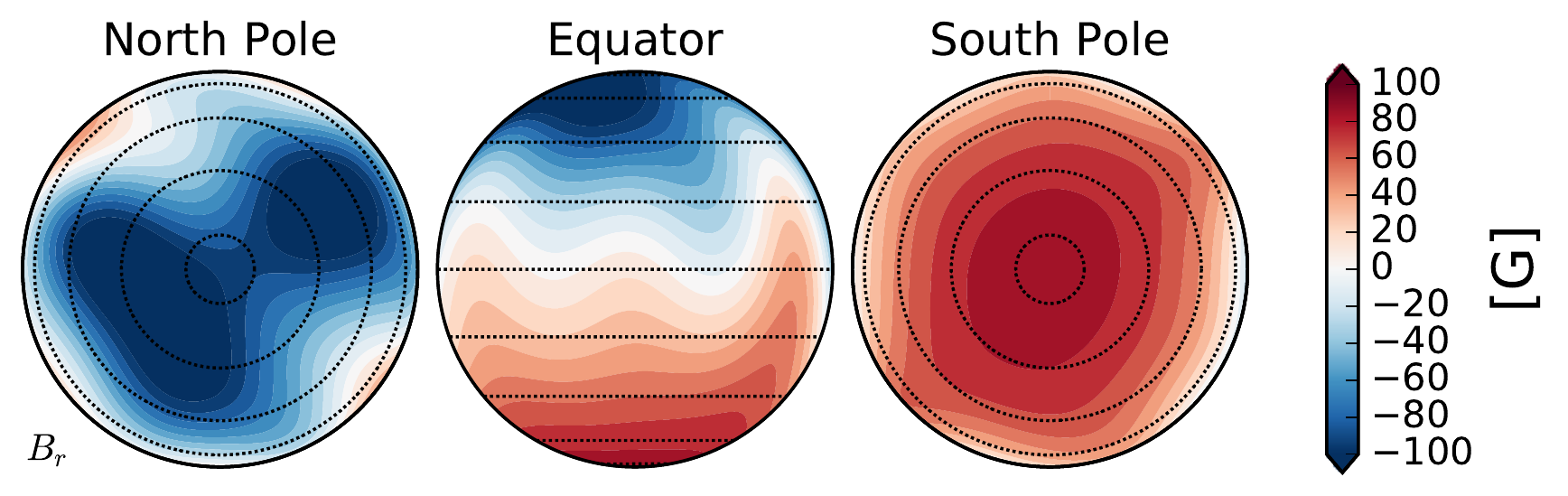} \\
\textbf{BD- 16351} & \textbf{TYC 5164-567-1} \\
\includegraphics[scale=0.5]{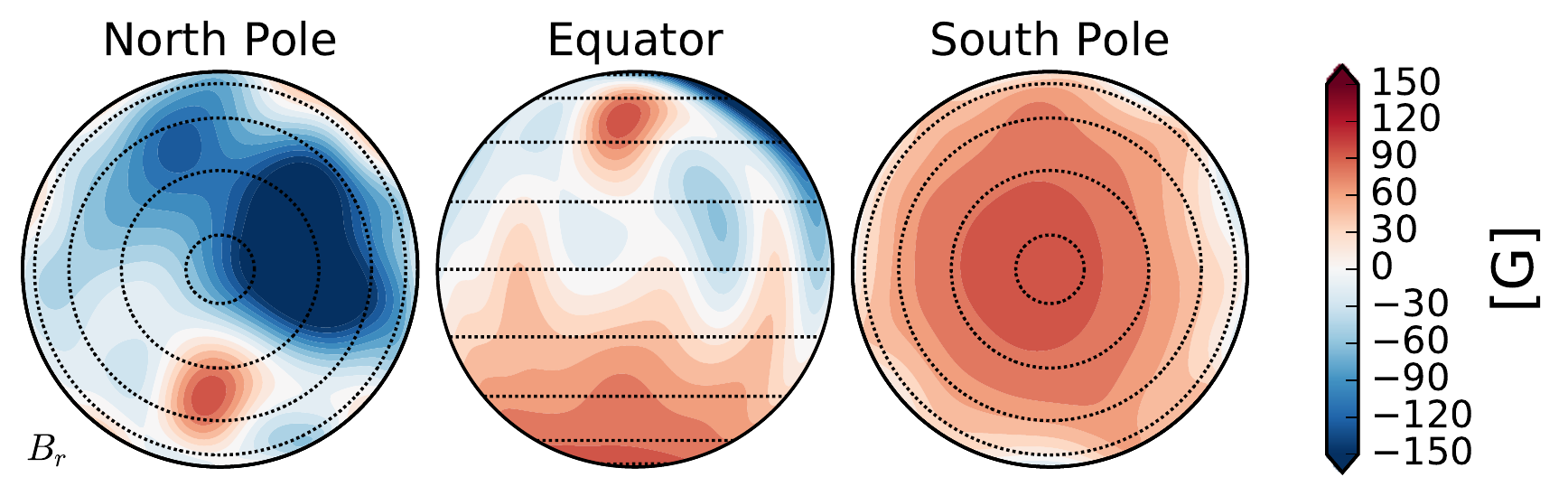} & \includegraphics[scale=0.5]{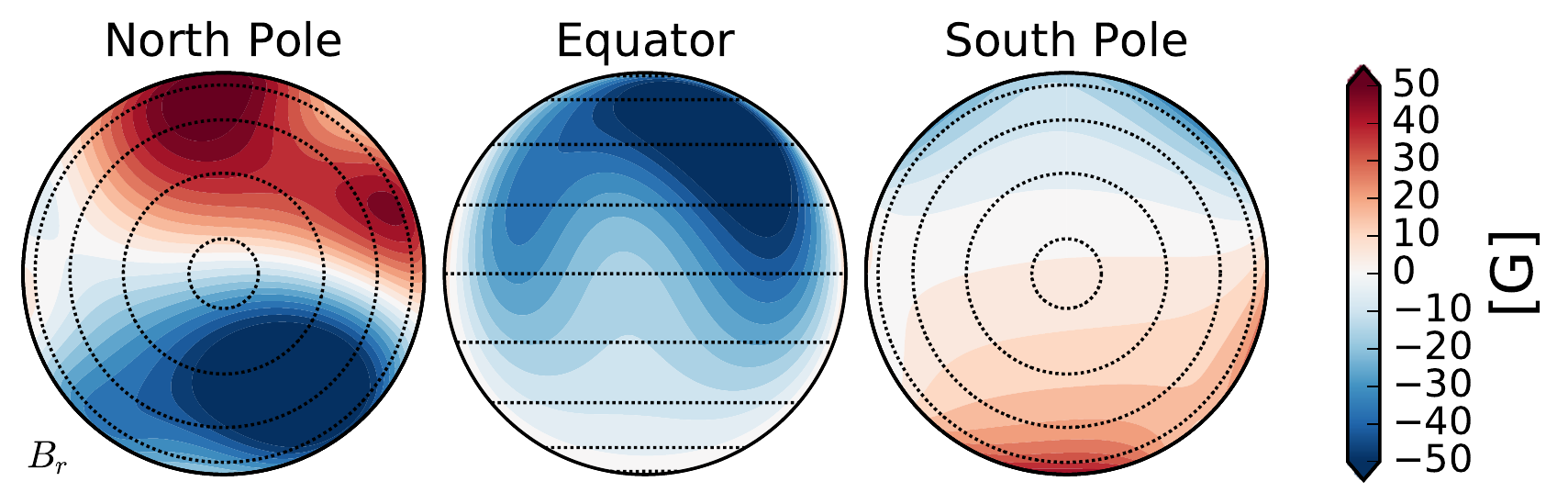} \\
\textbf{HII 296} & \textbf{DX Leo} \\
\includegraphics[scale=0.5]{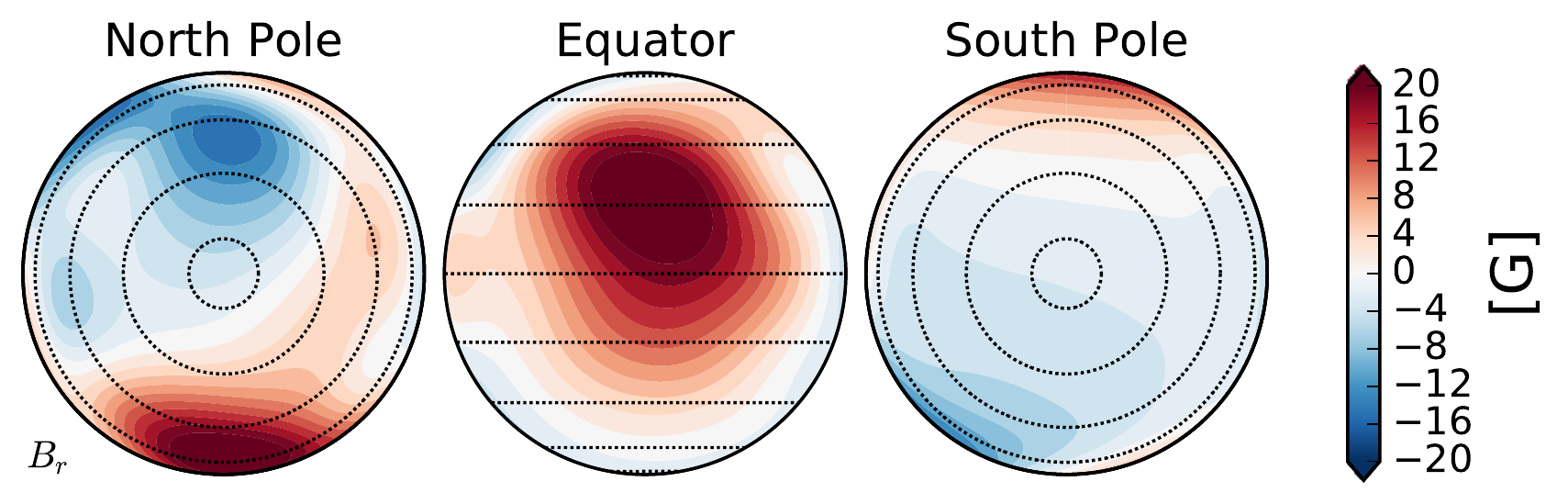} & \includegraphics[scale=0.5]{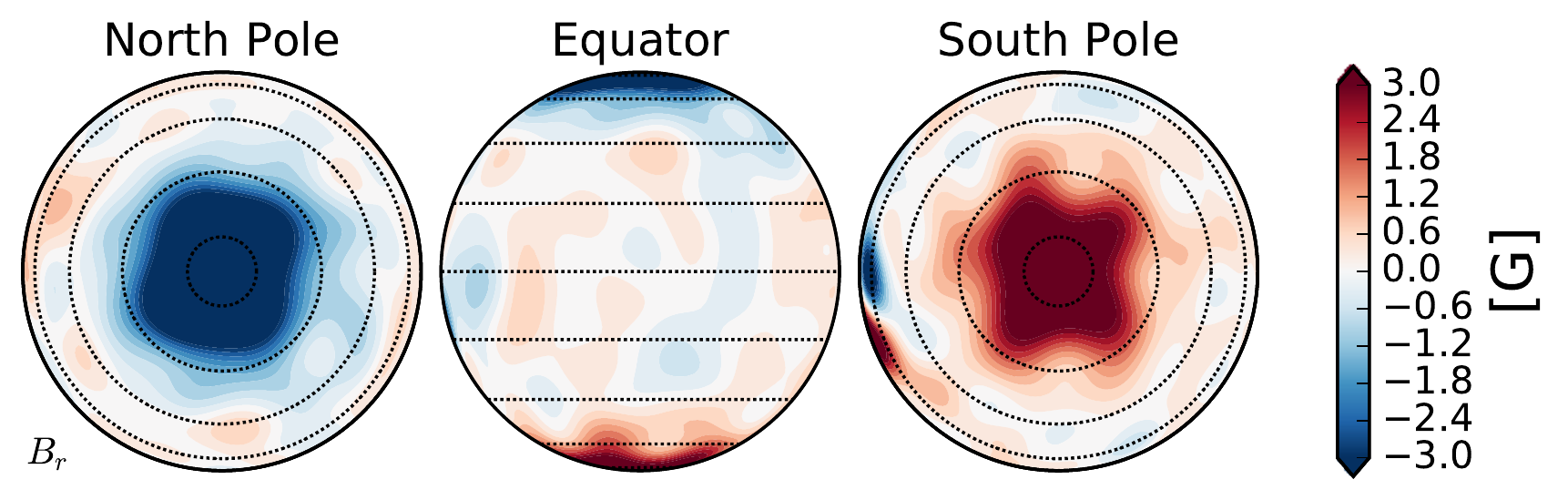} \\
\textbf{AV 2177} & \textbf{Sun Wilcox 1996}
\end{tabular}
\caption{Reconstruction of the radial magnetic field for each star. The saturation value of the color scale changes with the amplitude of the radial magnetic field of each case. For all stars, we observe a dominant dipole field, with smaller scale features. The overall amplitude of the field decreases with the rotation rate of the star, and maps are sorted by age.}
\label{zdi}
\end{figure*}

We consider in this paper six different stars whose ages are precisely determined. Five of those stars belong to the study made by \citet{Folsom2016} and share a mass and radius close to $(0.9 M_{\odot},0.9 R_{\odot})$. Their rotational periods and surface magnetic fields have been determined using observations from the spectropolarimeters NARVAL (operating at T\'elescope Bernard Lyot, France) and ESPaDOnS (operating at the Canada France Hawaii Telescope, Hawaii). Their ages are determined by studies of the open clusters and associations they belong to. They were chosen to span a range of ages and rotation rates but have similar masses, and bright stars were selected to have sufficient signal to noise ratio.

The sixth star we include in our study is the Sun, for which we used a magnetogram obtained at the Wilcox observatory \citep{DeRosa2012}, and which will serve as a reference and oldest star in this paper. It is well known that the solar magnetic topology of the Sun and the solar wind structure vary over the $11$-yr cycle \citep[see, e.g.,][]{Sokol2015,PintoBrun2011}. We considered the Sun in its minimum of activity in late 1996, during cycle 22. The ZDI maps of the five young stars are able to describe the surface magnetic fields of the stars up to a degree $\ell_{\mathrm{max}}=15$ \footnote{The actual resolution of ZDI is somewhat lower than this ($\ell_{\mathrm{max}} =8-10$), but the fields are reconstructed using $15$ spherical harmonics.} in a spherical harmonics decomposition, which represents a sum of $135$ different modes. The solar magnetograms made at the Wilcox observatory are able to reach much higher resolution ($\ell_{\mathrm{max}}=50$). We chose to cut the solar map to $\ell_{\mathrm{max}}=15$ to keep the same resolution for all stars in our sample. This partly justifies the choice of the solar minimum, whose energy spectrum is more concentrated in large scale structure than during maximum of activity. The first $15$ spherical harmonics represent $95\%$ of the magnetic energy during the minimum, and $80\%$ during the maximum for cycle 22. Although the ages are sampled logarithmically, the sampling of the rotation rates is enough to follow closely the changes in coronal parameters and in magnetic field amplitude (see Section \ref{subsec:evol}). The stellar parameters of all the targets are listed in Table \ref{table1}.

Figure \ref{zdi} shows the surface radial magnetic field reconstructed from the spherical harmonics decomposition of the Zeeman Doppler analysis \citep{Folsom2016}. The field is presented as orthographic projections on three different angles, with views facing the equator and the two poles. The color scales are chosen according to the amplitude of the magnetic field of the star. The average radial magnetic field of each case is given in table \ref{table1} and mostly increases with the rotational frequency, as expected by dynamo theory \citep{DurneyLatour1978,Weiss1994,Brun2015}. We can see that, for each case, a dominant dipolar component is present, alongside smaller scale modes. This also motivated our choice of the solar minimum of activity, which exhibits a mostly dipolar field. Nonetheless, the dipolar components can show large inclinations with respect to the rotation axis, thus being far from an axisymmetric configuration and making a 3D approach necessary.

These magnetic maps are used as boundary conditions and specify the surface magnetic field of our simulations. However, they do not properly describe the stellar parameters by themselves. Like the magnetic field amplitude and topology, the thermodynamical properties of the base of the corona are likely to change with age and rotation \citep{Gudel2004,Giampapa2005}. The next section is dedicated to describing the model we used to take into account those variations.

\subsection{Evolution of coronal properties with age}
\label{subsec:evol}

Our numerical model needs, in addition to the surface magnetic field, assumptions for the coronal base temperature and density. Several studies have addressed the evolution of those parameters with age and other stellar properties \citep{HolzwarthJardine2007,CranmerSaar2011,Suzuki2013b}. Among others, their objective was to explain the mass loss rate signatures in the astrospheres' Ly$\alpha$ absorption spectra \citep[see][]{WoodRev2004,Wood2005a}. Those studies are also nourished with a long history of X-ray coronal emissions \citep{Pallavicini1981,Pallavicini1989,Pizzolato2003,Gudel2004,Wright2011}, which show that coronal densities and temperatures tend to increase with the rotational period in solar-like stars. For instance, \citet{HolzwarthJardine2007} gave scaling laws for the evolution of the temperature $T$ and number density $n$ as a function of the rotational frequency $\Omega_{\star}$ only assuming a power law dependence:

\begin{equation}
T = T_{\odot} \left( \frac{\Omega_{\star}}{\Omega_{\odot}} \right)^{0.1},\quad n = n_{\odot} \left( \frac{\Omega_{\star}}{\Omega_{\odot}} \right)^{0.6}.
\end{equation}

The values of the exponent for the power law correspond to their reference case, which aims to match the lowest branch of mass loss measurements. In their model, the mass loss is obtained computing 1D polytropic and magneto-centrifugal wind from the coronal parameters \citep[see][]{WeberDavis1967,Sakurai1985,Reville2015b}. The study of \citet{Wood2005a} shows, however, mass loss rates that can reach $100$ times the solar value for rather slow rotators \citep[see the case of $70$ Oph, with a period of $\approx 20$ d,][]{Wood2005a}. Those extreme values need a stronger increase of the density with the rotation rate. \citet{Suzuki2013b}, using simulations of flux tubes heated by Alfv\'en wave dissipation, showed that such values could be reached and that a saturation could be obtained through the increase of the coronal density that increases the radiative losses. The dependence of the coronal density on $\Omega_{\star}$ is, however, also constrained by the observed X-ray fluxes and spin evolution models that suggest that $0.6$ is a good estimate for the exponent \citep{IvanovaTaam2003}.

Moreover, the reference case fits the supposedly weak mass loss of more active stars (corresponding to most of the ages and rotation rates selected in our study) without invoking an additional transition below the $F_X$ saturation threshold, which still requires further theoretical understanding \citep{Vidotto2016a}. We thus chose to keep the same prescriptions as \citet{HolzwarthJardine2007}, changing the solar reference temperature and density, as we use a different value for the polytropic index $\gamma$. In our model, $T_{\odot}=1.5\times10^{6}$ K and $n_{\odot}=10^{8}$ cm$^{-3}$ are calibrated such that a wind with $\gamma=1.05$ recovers a wind velocity of $444$ km s$^{-1}$ at 1 AU and a mass loss rate of $3.2 \times 10^{-14} M_{\odot}/$yr. This value for $\gamma$ has been widely used in the literature, including our works \citep{WashShib1993,Matt2012,Reville2015a}, to describe the quasi-isothermal expansion of the wind with a polytropic model.

The temperature thus varies from $1.5 \times 10^{6}$ K to $1.9 \times 10^{6}$ K and the density from $1 \times 10^8$ cm$^{-3}$ to $4.2 \times 10^{8}$ cm$^{-3}$ throughout our sample (see table \ref{table2}). 

\subsection{Computational methods and boundary conditions}
\label{subsec:num}

In this study, we numerically solve the time-dependent ideal magnetohydrodynamics equations until a steady state is reached in our wind simulations. We use the PLUTO code \citep{Mignone2007}, using a finite-volume Godunov type scheme and a Harten, Lax, van Leer, and Einfeldt (HLLE) solver \citep{Einfeldt1988} in three dimensions. Finite volume methods aim to provide fully compressible and shock capturing numerical methods that consider fluxes of conservative quantities through volumes. Hence, they formulate the MHD equations as a set of eight conservation equations defined as follows:

\begin{equation}
\label{MHD_1}
\frac{\partial}{\partial t} \rho + \nabla \cdot \rho \mathbf{v} = 0,
\end{equation}
\begin{equation}
\label{MHD_2}
\frac{\partial}{\partial t} \mathbf{m} + \nabla \cdot (\mathbf{mv}-\mathbf{BB}+\mathbf{I}p) = - \rho \nabla \Phi + \rho \mathbf{a},
\end{equation}
\begin{equation}
\label{MHD_3}
\frac{\partial}{\partial t} (E + \rho \Phi)  + \nabla \cdot ((E+p+\rho \Phi)\mathbf{v}-\mathbf{B}(\mathbf{v} \cdot \mathbf{B})) = \mathbf{m} \cdot \mathbf{a},
\end{equation}
\begin{equation}
\label{MHD_4}
\frac{\partial}{\partial t} \mathbf{B} + \nabla \cdot (\mathbf{vB}-\mathbf{Bv})=0,
\end{equation}
where the energy $E \equiv  \rho \epsilon + \rho v^2/2 + B^2/2$, the magnetic field $\mathbf{B}$, the mass density $\rho$, and the momentum  $\mathbf{m} \equiv \rho \mathbf{v}$ are the conservative variables. Here, $\mathbf{v}$ is the  velocity field, $p = p_{th} +B^2/2$ is the total (thermal plus magnetic) pressure and $\mathbf{I}$ is the identity matrix. The potential $\Phi$ accounts for the gravitational attraction of the star and $\mathbf{a}$ is a source term that contains the Coriolis and centrifugal forces as we solve the equations in a rotating frame. The magnetic field is split into a background and a variable component for computational purposes \citep[see][]{Powell1994}. An ideal equation of state is used to close the set of MHD equations, and the internal energy is written

\begin{equation}
\epsilon = \frac{p}{\rho(\gamma-1)},
\end{equation}

with $\gamma=1.05$, the ratio of specific heats, which differs from the usual value of $5/3$ for a hydrogen gas in order to mimic the extended coronal heating.

We solve the equation in a cartesian geometry with a grid centered on the star that extends from $-30$ to $30$ stellar radii in each direction. The grid is uniform in a cube of $[-1.5 R_{\star}, 1.5 R_{\star}]^3$ with $192$ grid points in each direction and then stretched up to $30 R_{\star}$ with an additional 256 points for each direction. The resolution at the surface of the star is $50\%$ larger than the one used in \citet{Reville2015a}, but several tests have demonstrated that this resolution is enough to have numerical convergence.

The initialization is done by setting a spherically symmetric profile of a $\gamma=1.05$ polytropic wind for the density, pressure and poloidal velocity. This initial solution is obtained by a Newton-Raphson method on the normalized velocity and the critical radius of the polytropic wind solution. The MHD equations are then solved in a frame rotating with the star. We only initialize a solid body rotation inside the star so that it is the magnetic field that gives its rotating motion to the surrounding plasma. The magnetic field is initialized with a potential field extrapolation \citep{Schatten1969} using the radial component of the ZDI map and a source surface radius $r_{\mathrm{ss}} = 15 R_{\star}$. This particular initial choice of the source surface radius has no impact on the final state since the extrapolated potential field then dynamically evolves with the stellar winds toward a relaxed state.

Boundary conditions at the surface of the star -which model the base of the corona in our case \footnote{See \citet{MatsumotoSuzuki2012} for a 2D model of the chromosphere and the transition region.}- are a sensitive point of our study. As in \citet{Reville2015a}, we set three layers with different properties. For all layers, density and pressure are maintained as the initial transonic polytropic wind solution. In the top layer, the poloidal velocity is set to be parallel to the magnetic field, while the toroidal velocity and the magnetic field are free to evolve. In the middle layer, the magnetic field is still free, the poloidal velocity is zero and the toroidal velocity is set for solid body rotation. Finally, in the deepest layer, we enforce the reconstructed magnetic field, considering a perturbation in the toroidal field that self adapts to minimize the overall current. More explicitly, the magnetic field solution interacting with the rotating wind in open regions will be in general different from the potential extrapolation we set at the initialization. Hence, to ensure a current free magnetic field inside the star, which is supposed to be a perfect conductor, we dynamically modify $B_{\varphi}$ in the deepest layer to get as close as possible to a curl free magnetic field in the rotating frame \citep[see][]{MattBalick2004,ZanniFerreira2009}. This boundary condition has a strong effect on the conservation of MHD invariants. For instance, the quantity that corresponds to the derivative of the electric field potential in axisymmetric configurations \citep[see][]{Reville2015a,Matt2012,KG2000,Ustyugova1999} remains in our 3D simulations close to the stellar rotation rate when our boundary condition is applied. We will see in subsection \ref{subsec:semianal} that this condition is key for an improved treatment of the angular momentum loss computation.

The outer edges of our domain are treated with \textit{outflow} boundary conditions that set the derivative of each field normal to the boundary to zero.

\section{Results: Global Properties}
\label{sec:glob}

\subsection{Mass and angular momentum loss}

\begin{figure*}
\center
\begin{tabular}{rcr}
\includegraphics[scale=0.4]{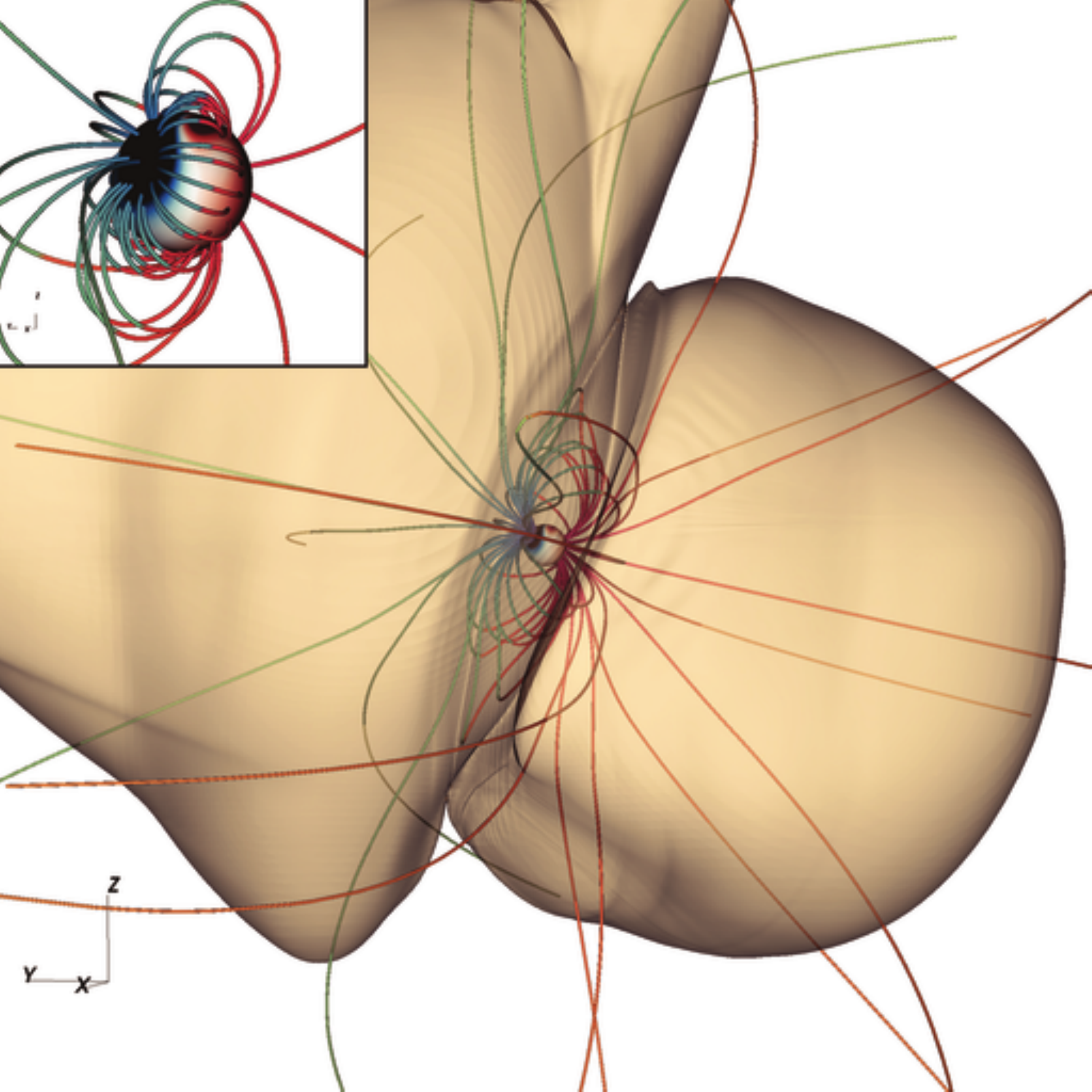} & \hspace{1cm}  &\includegraphics[scale=0.4]{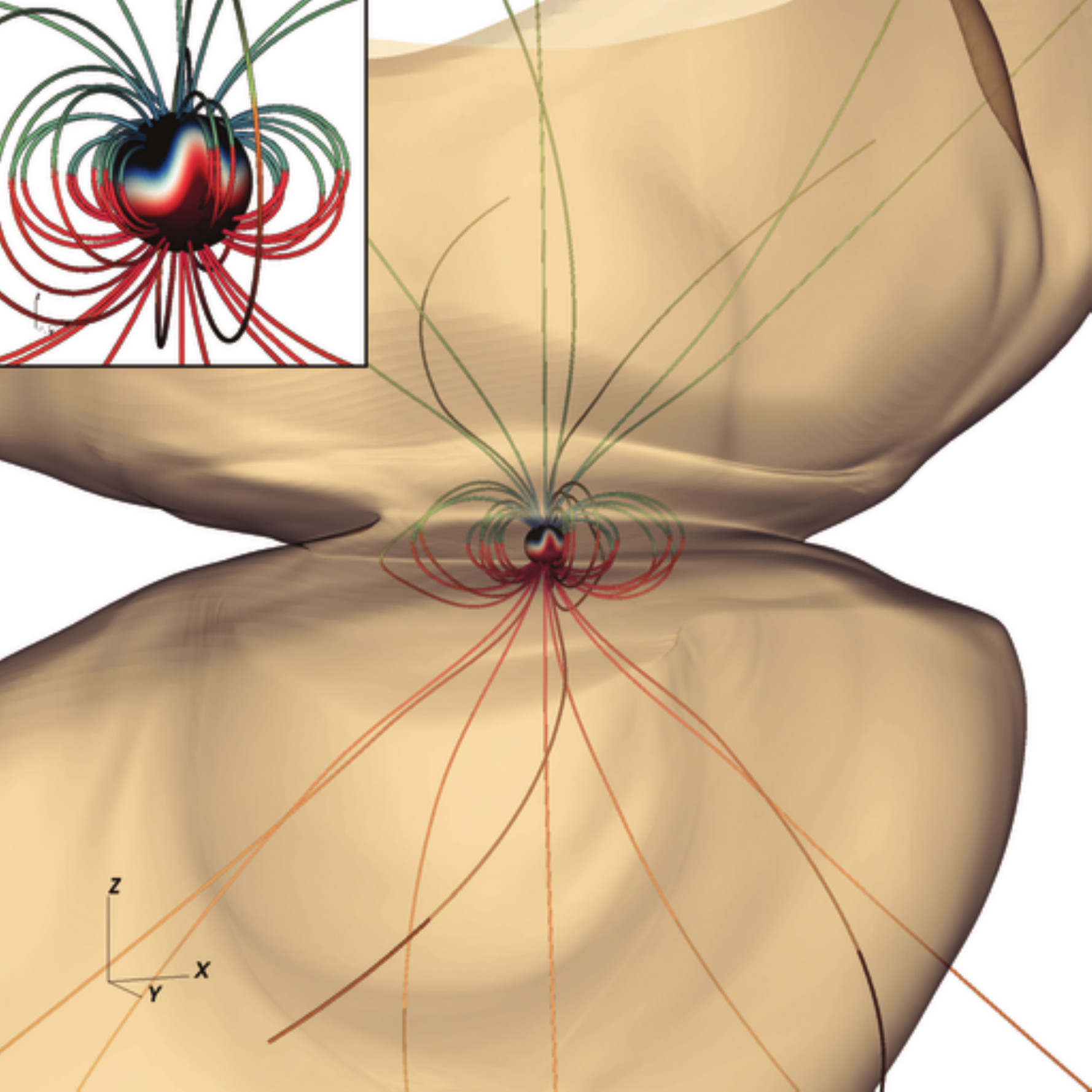} \\
\textbf{BD- 16351} & \hspace{1cm} & \textbf{TYC 5164-567-1} \\
\includegraphics[scale=0.4]{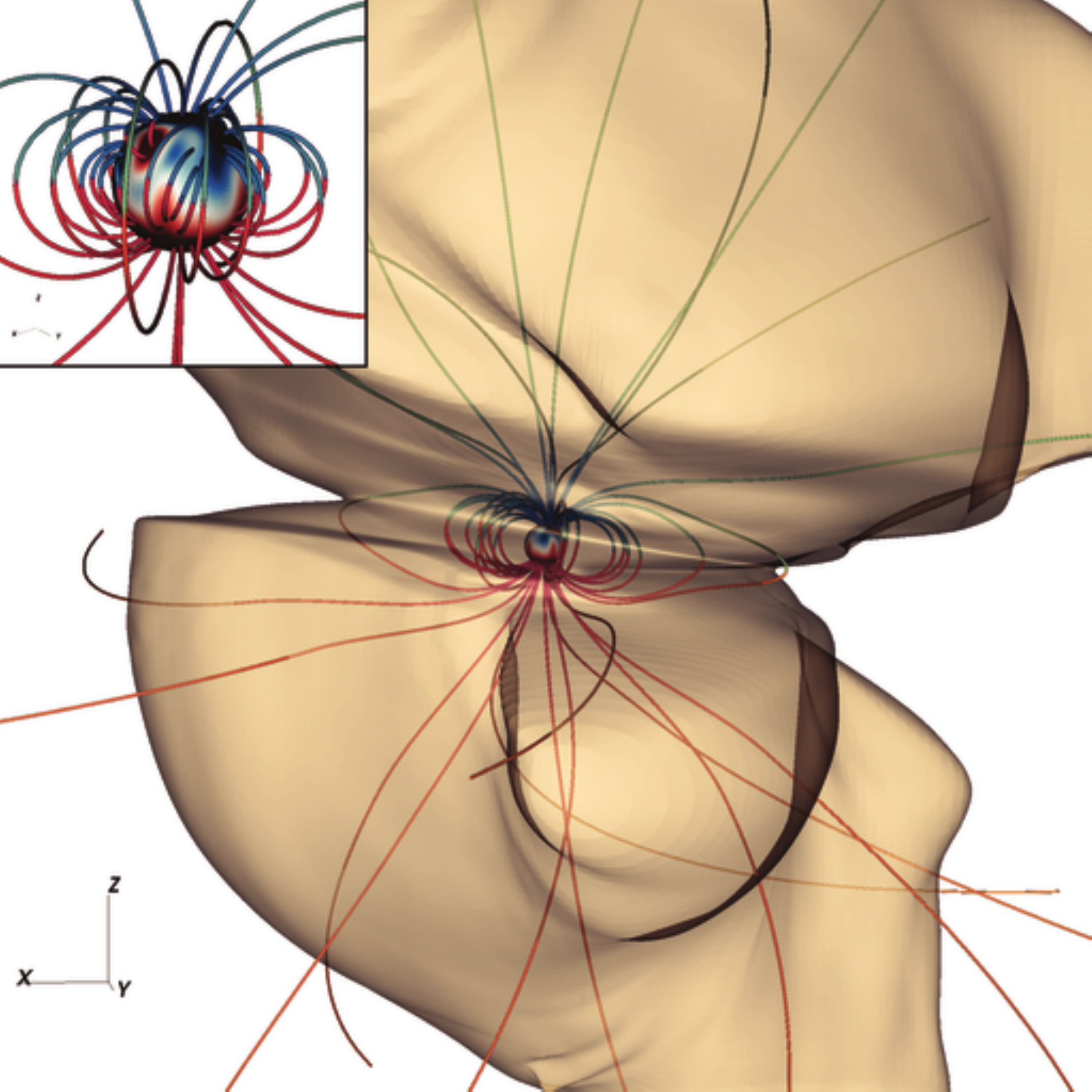} & \hspace{1cm} &\includegraphics[scale=0.4]{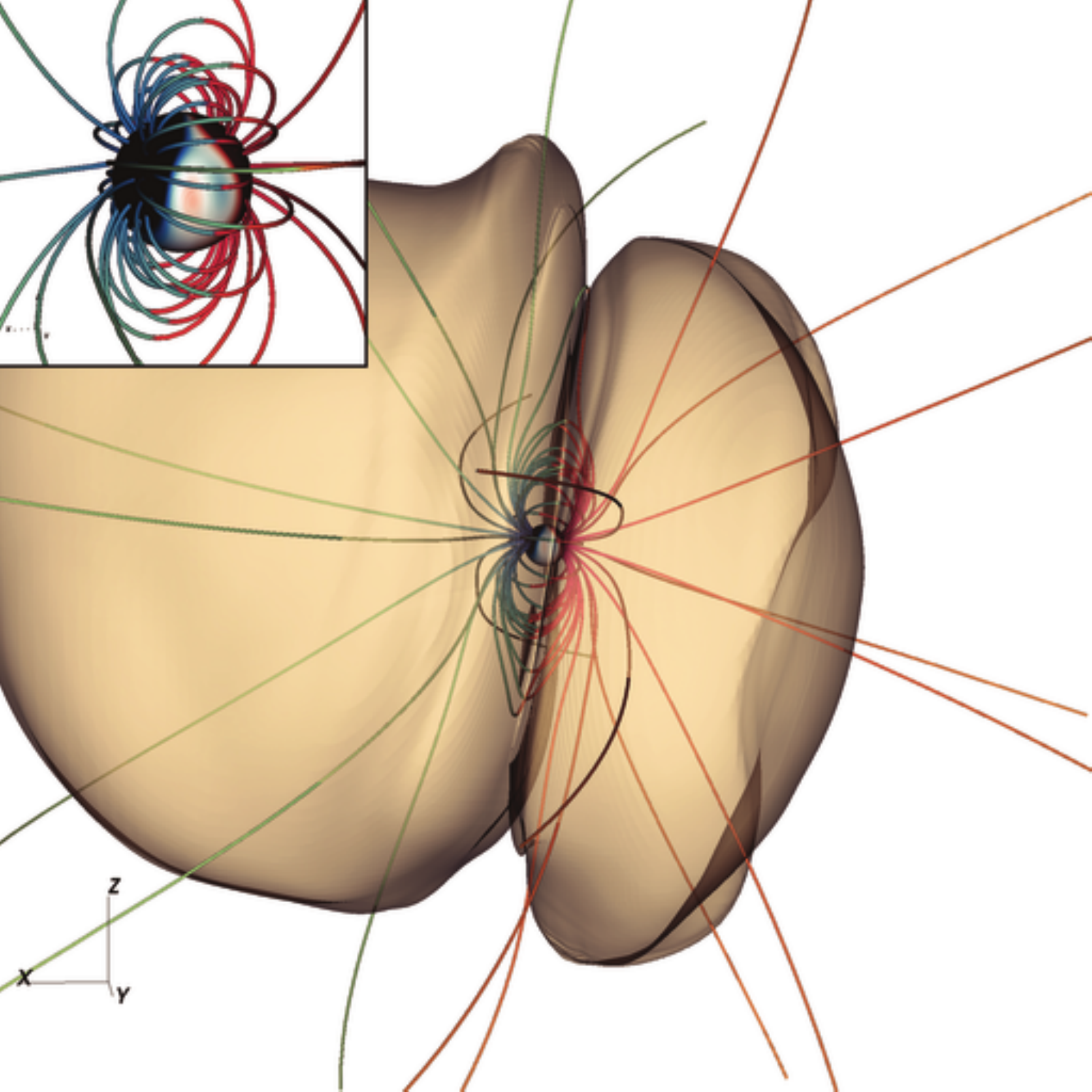} \\
\textbf{HII 296}   & \hspace{1cm} & \textbf{DX Leo}\\
\includegraphics[scale=0.4]{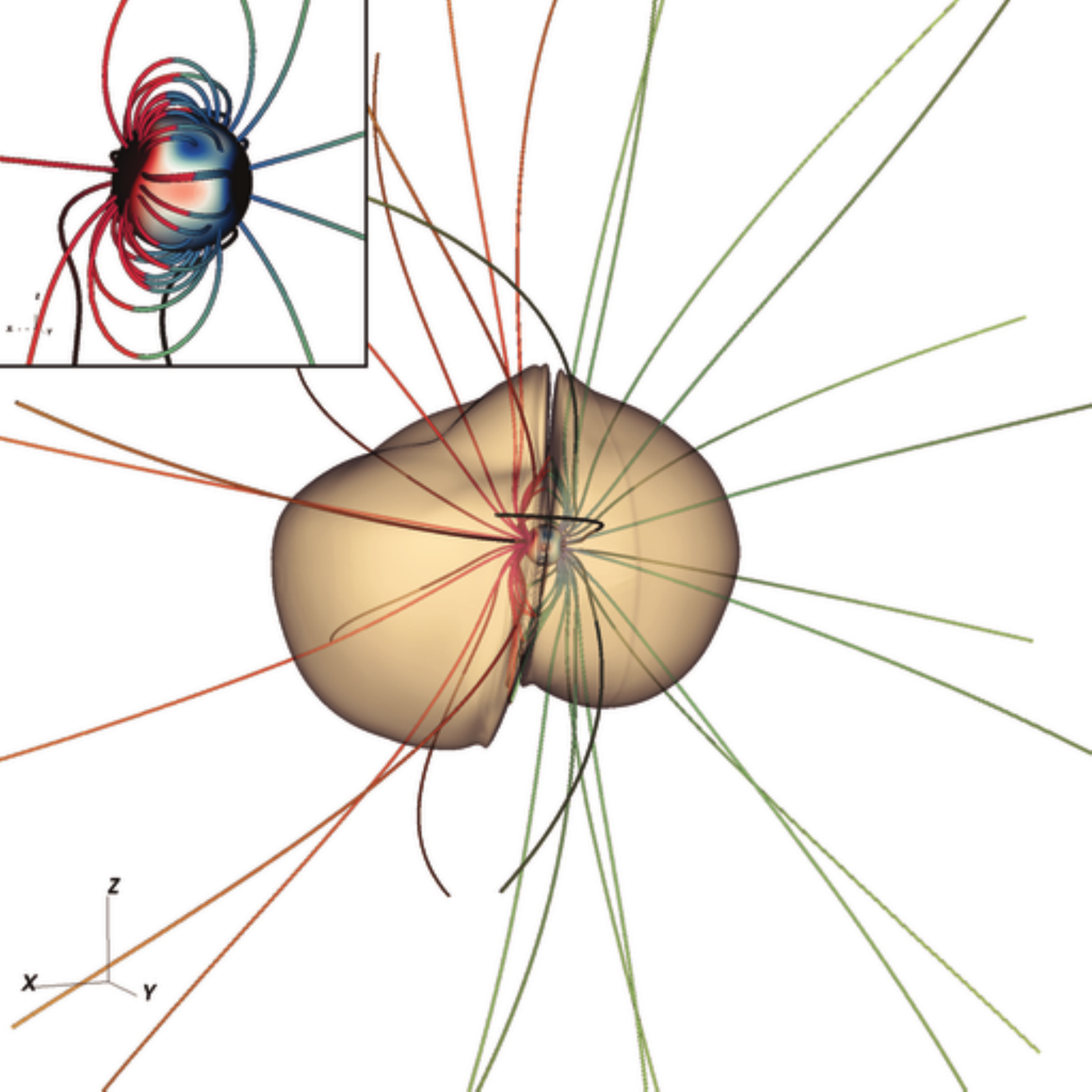}  & \hspace{1cm} & \includegraphics[scale=0.4]{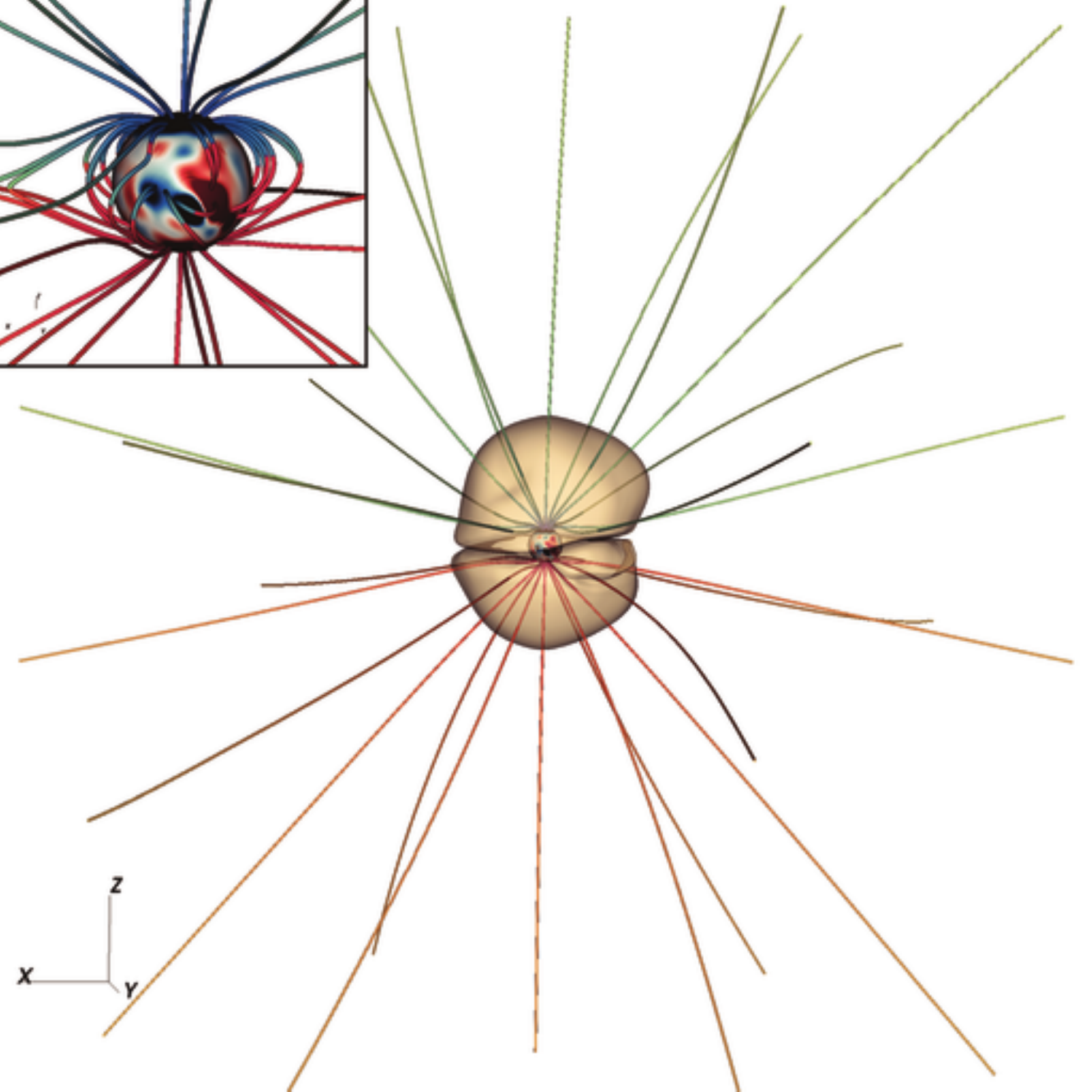}\\
\textbf{AV 2177}  & \hspace{1cm} & \textbf{Sun 1996} 
\end{tabular}
\caption{Steady state of the 3D simulations. The Alfv\'en surfaces are shown in beige. The surface radial field is shown at the surface of the star and field lines of positive polarity are shown in red, negative polarity in blue. The Alfv\'en surfaces are largely shaped by the dominant dipole of each case. However, they are irregular due to the dipole inclination and higher order components of the magnetic field. The size of the Alfv\'en surface grows with the amplitude of the field. We show a zoom-in on the stellar surface with smaller scale features at the surface in the top left corner of each panel.}
\label{3dim}
\end{figure*}

Figure \ref{3dim} shows the resulting steady-state wind solutions achieved in the six simulations. The convergence time is typically of the order of a few times the Alfv\'en time scale, \textit{i.e.} the time for alfv\'en waves to cross the simulation's domain. Given the high resolution, and the time step that can vary by one order of magnitude depending on the magnetic field amplitude, each of these simulations reaches a steady states after $10^5$ to $2 \times 10^6$ times steps (between $10^5$ to $5 \times 10^5$ core hours on supercomputers).

As usual, the topology of the coronal magnetic field in steady state can be divided into open field regions, or coronal holes, and closed field regions, or dead zones, where the plasma corotates with the star. The surface magnetic field of all the stars in our sample includes a significant dipole, that can be strongly inclined. This mode gives its large scale structure to the coronal magnetic field. However, the magnetic field inclination, amplitude and smaller scale components lead to irregular shapes of the Alfv\'en surface, which is shown in Figure \ref{3dim}. For some cases, the Alfv\'en surface extends beyond the simulation domain. Indeed, for fast rotators, field collimation induces an increase of the poloidal magnetic field amplitude near the rotation axis, and the Alfv\'en surface is pushed farther away. A precise description of this phenomenon can be found in \citet{WashShib1993,Ferreira2013,Reville2015a} and we will address some of its consequences in section \ref{sec:vel}. However, the global properties we are interested in, such as the mass and angular momentum loss rates are constants within a small numerical variation once integrated over a surface that encloses the largest closed coronal loops. In the case of TYC 5164-561-1, which seems to have a significant part of its Alfv\'en surface out of the computation domain (more than any other case), this numerical variation of the mass and angular momentum loss is below 3\%. It drops below 1\% for cases where the Alfv\'en surface is fully inside the computation domain.

The mass loss rate $\dot{M}$ and the angular momentum loss rate $\dot{J}$ associated with the wind are computed as

\begin{equation}
\dot{J} = \int_S \rho \Lambda \mathbf{v} \cdot d\mathbf{S},
\label{Jdot}
\end{equation}
where

\begin{equation}
\Lambda = R \left( v_{\varphi} - B_{\varphi}\frac{B_p}{\rho v_p} \right),
\label{Lambda}
\end{equation}
and
 
\begin{equation}
\dot{M} = \int_S \rho \mathbf{v} \cdot d\mathbf{S}.
\label{Mdot}
\end{equation}

The subscript $p$  and $\varphi$ stand for the poloidal and azimuthal components of each vectorial field, and $R$ is the cylindrical radius. Those integrals can be computed from any surface $S$ that contains all the closed coronal loops. From those outputs we define an effective Alfv\'en radius, which conveniently matches the relation given in the simplified model of \citet{WeberDavis1967}:

\begin{equation}
\langle R_A \rangle = \sqrt{\frac{\dot{J}}{\Omega_{\star} \dot{M}}}.
\label{AvAlf}
\end{equation}

This effective value matches quantitatively the average cylindrical radius on the irregularly shaped Alfv\'en surface of our simulations. All the global quantities computed from our simulations are given in Table \ref{table2}. We see that the angular momentum loss (AML) varies by a factor of $470$ from the Sun to the fastest rotator HII 296. The variation of the mass loss is lower, with values that reach $6$ times the solar value, here defined as $3.0 \times 10^{-14} M_{\odot}$/yr. The position of the Alfv\'en radius is globally increasing with rotation rate but is also very sensitive to temperature. For instance, the largest value we get is $16.6 R_{\star}$ for TYC 5164-567-1, which has a similar magnetic field to HII 296, but a slightly cooler coronal temperature, which makes the wind slower in our model.

One can recover the convergence on the Skumanich law for solar-like stars assuming a loss of angular momentum proportional to $\Omega_{\star}^3$ in the non-saturated regime. In the saturated regime, although there is no consensus\footnote{For instance, a purely spherically symmetric radial Weber and Davis model yields a self-consistent saturation where $\dot{J} \propto \Omega_{\star}^2$ \citep[see][]{Keppens1995}.}, the dependence of the angular momentum loss is usually assumed to be linear with $\Omega_{\star}$. Hence, a canonical expression for the stellar wind torque can be written as \citep{Kawaler1988,Bouvier1997}\footnote{Recent work of \citet{Matt2015} proposed a more accurate description of the mass dependence.}:

\begin{equation}
\dot{J} = \dot{J}_{\odot} \left( \frac{\mathrm{min}(\Omega_{\star},\Omega_{\mathrm{sat}})}{\Omega_{\odot}} \right)^{2} \left(\frac{\Omega_{\star}}{\Omega_{\odot}} \right) \left( \frac{M_{\odot}}{M_{\star}}\frac{R_{\star}}{R_{\odot}} \right)^{0.5}
\label{EmpiricalModel}
\end{equation}
where $\Omega_{\mathrm{sat}}$ is a saturation value of the angular momentum loss, which occurs at $\Omega_{\mathrm{sat}} = 8 \Omega_{\odot}$ for K-type stars. This saturation value corresponds to a Rossby number $Ro \approx 0.1$ \citep{Wright2011}, and higher mass stars have higher $\Omega_{\mathrm{sat}}$ \citep[for G stars the value is around $15 \Omega_{\odot}$, see][]{GalletBouvier2015}. For clarity, we define $\tilde{\Omega}_{\star}= \mathrm{min} (\Omega_{\star},\Omega_{\mathrm{sat}})$, with $\Omega_{\mathrm{sat}} = 8 \Omega_{\mathrm{sat}}$. Moreover, we have for every star in our sample $f(M_{\star},R_{\star}) \equiv \left(M_{\star}/M_{\odot}\right)^{-0.5}  \left(R_{\star}/R_{\odot}\right)^{0.5} \approx 1$ so that formulation (\ref{EmpiricalModel}) reduces to $\dot{J}/\dot{J}_{\odot} = \tilde{\Omega}_{\star}^2 \Omega_{\star}/\Omega_{\odot}^3$.

\begin{figure}
\center
\includegraphics[scale=0.45]{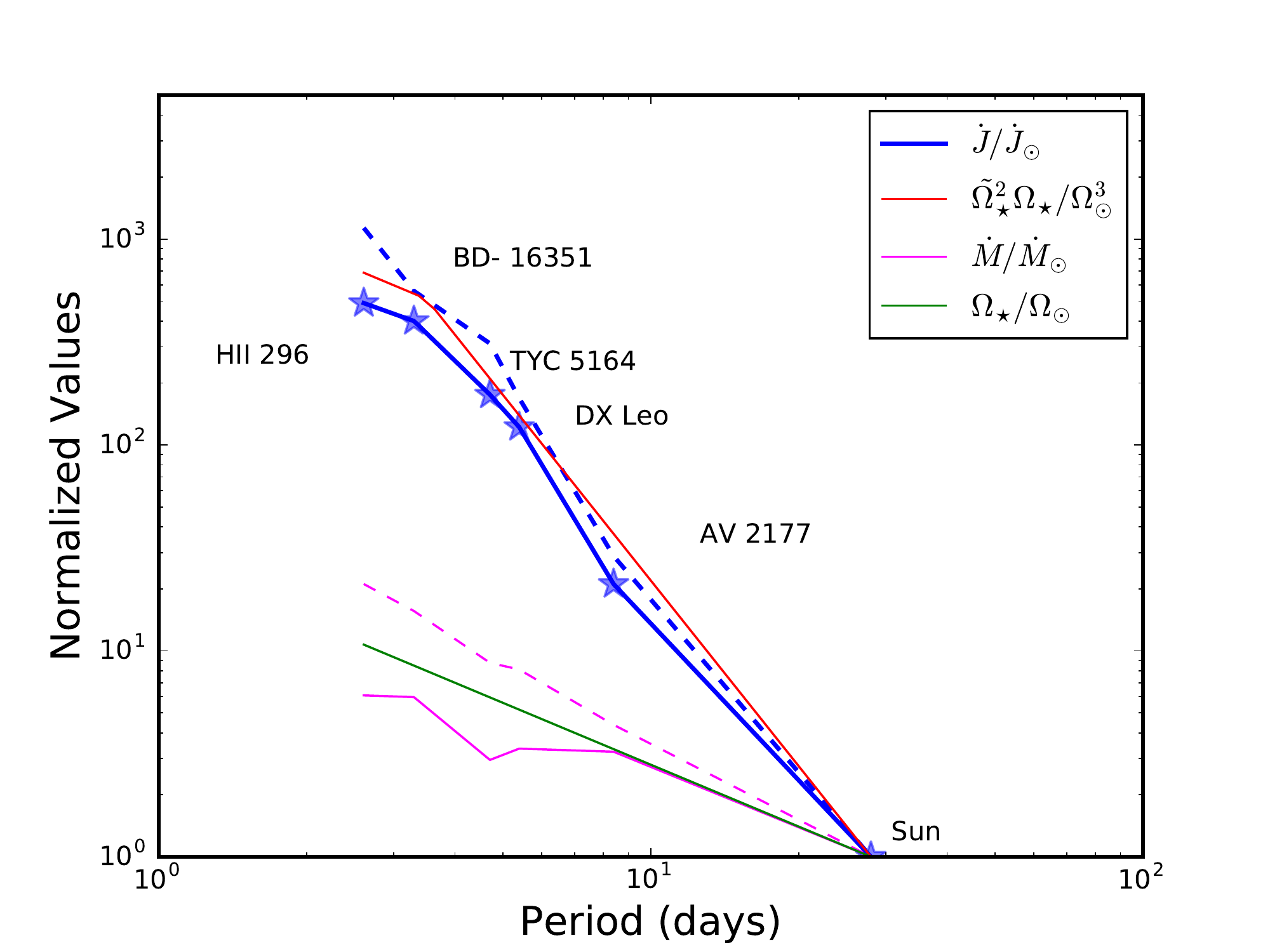}
\caption{Evolution of the mass and angular momentum loss with age, compared with the rotational frequency and equation (\ref{EmpiricalModel}). The angular momentum loss in blue follows the empirical law $\dot{J} \propto \tilde{\Omega}_{\star}^2 \Omega_{\star}$. The variation of the mass loss is shown in magenta. The variation of the mass loss is close to linear with respect to $\Omega_{\star}$ and is thus not enough to explain the total AML variation, which needs an increase of the Alfv\'en radius. The results of the semi-analytical model are added with dashed lines.}
\label{OmegaDep}
\end{figure}

In Figure \ref{OmegaDep}, we compare the resulting torque from our simulations (in blue) to this empirical formulation (\ref{EmpiricalModel}) (in red). All quantities have been normalized by the solar value. We look only at the $\Omega_{\star}$-dependence of the angular momentum loss from the simulation, and the agreement is good. This agreement is the result of the combination of the observed magnetic fields, the hypothesis on the coronal temperature and density and the simulation methods. Taking into account complex, three dimensional magnetic fields is necessary, as some of our targets have a significant part of their magnetic energy in non-axisymmetric modes. Moreover, we can see an inflection in the slope for the two fastest rotators BD- 16351 and HII 296, which corresponds to the saturation value (which is not used whatsoever in our simulation). The saturation seems to appear self-consistently in our simulations, thanks to the plateau reached by the mass loss rates for our fast rotators (we will come back to this point in the next section).

\begin{deluxetable*}{r|c|c|c|c|c|c|c|r}
  \tablecaption{Input parameters and results of the simulations}
  \tablecolumns{9}
  \tabletypesize{\scriptsize}
  \tablehead{
    \colhead{Name} &
    \colhead{$\Omega_{\star}/\Omega_{\odot}$} &
    \colhead{$n$ ($10^{8}$ cm$^{-3}$)} &
    \colhead{$T$ ($10^6$ K)} &
    \colhead{$\langle R_A / R_{\star} \rangle$} &
    \colhead{$\dot{J} / \dot{J}_{\odot}$}&
    \colhead{$\dot{M} / \dot{M}_{\odot}$} &
    \colhead{$r_{\mathrm{ss,opt}}$} &
    \colhead{$r_{\mathrm{ss,est}}$}
  }
  \startdata
  BD- 16351 & 8.5 & 3.6 & 1.85 & 13.9 &380&5.5&8.1& 7.7 \\
  TYC 5164-567-1 & 5.9 & 2.9 & 1.8 & 16.7  &190&2.75&10.7& 10.5 \\
  HII 296 & 10.7 & 4.15 & 1.9 & 13.8 &470&5.65&9.3&8.7\\
  DX Leo & 5.2 & 2.7 & 1.76 & 13.3 &120&3.1&7.6& 7.4\\
  AV 2177 & 3.3 & 2.06 & 1.7 & 7.0 &20&3.0&4.6&4.3\\
  Sun & 1.0 &1.0 & 1.5 & 4.4 &1.0 & 1.0 &2.7&2.7\\
  \enddata
  \tablecomments{Evolution of the input parameters and results of the simulation with age and rotational frequency. The effective Alfv\'en radii are coherent with the Alfv\'en surfaces shown in Figure \ref{3dim}, with comparable sizes for the four youngest stars and much smaller and decreasing values for AV 2177 and the Sun. The optimal and estimated $r_{\mathrm{ss}}$ are very close to each other with slightly higher value for the optimal, although the change in resulting open flux is very small.}
  \label{table2}
\end{deluxetable*}

The mass loss varies in our simulation (magenta line) approximately linearly with $\Omega_{\star}$ (green line) and thus covers one order of magnitude over the sample. The variation of the angular momentum loss, which covers three orders of magnitude, is the result of three ingredients (see equation \ref{AvAlf}), the mass loss rate, the rotation rate and the average Alfv\'en radius squared, and we can say that each of these ingredients accounts for one order of magnitude. Actually, the Alfv\'en radius squared seems to account for a bit more than the mass loss rate, but those are not independent parameters, and this statement must be handled with caution.

\subsection{Semi-analytical model}
\label{subsec:semianal}

\begin{figure}
\center
\includegraphics[scale=0.45]{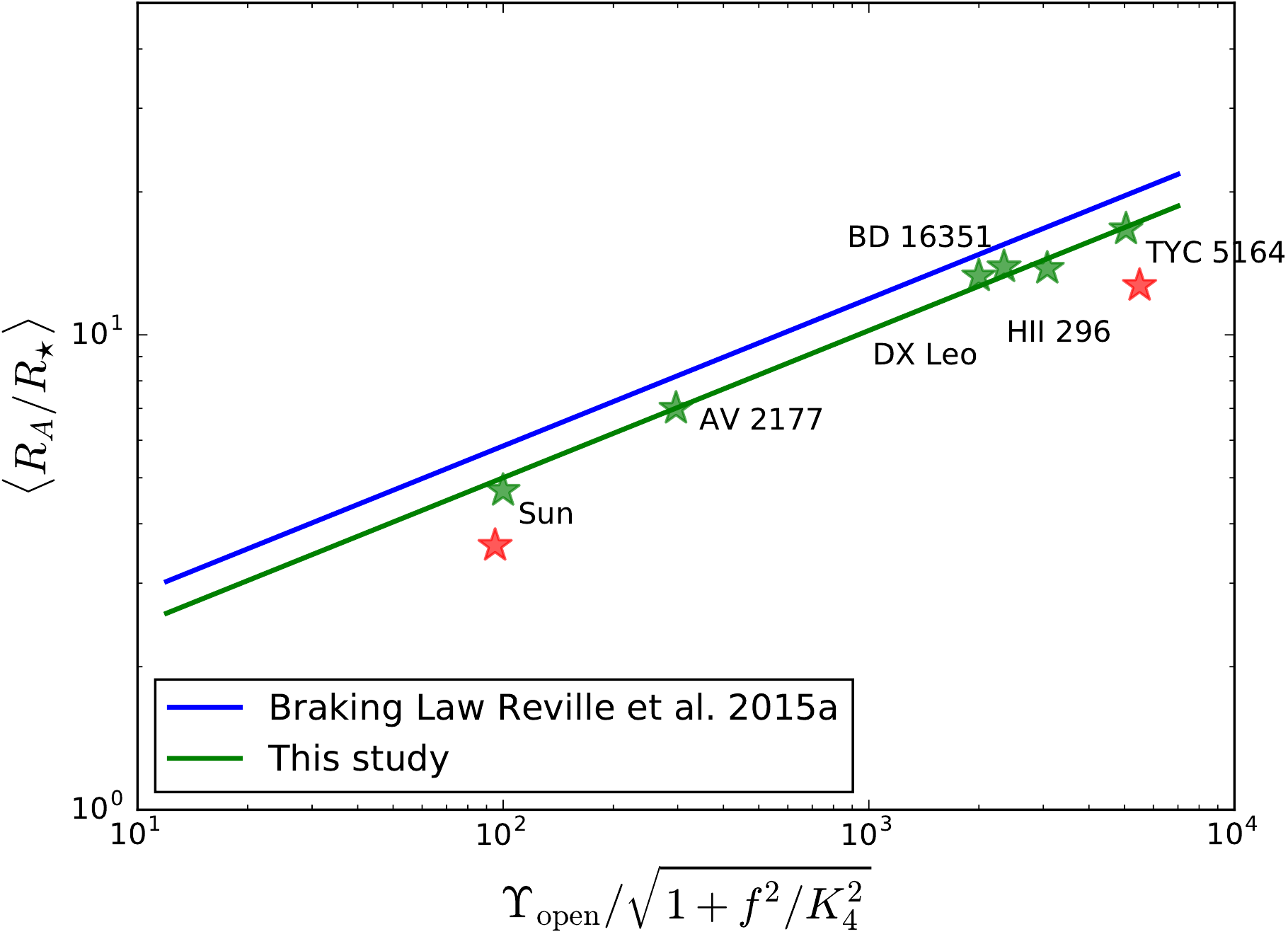}
\caption{Comparison of the 3D cases and the braking law of \citet{Reville2015a}. A new fit reducing by $15\%$ the constant $K_3$ is needed for a good agreement with the data, due to higher coronal temperatures in our sample (green stars). Red stars stand for simulations of the Sun and TYC 5164-567-1 with simpler boundary conditions. Those Alfv\'en radii are off the braking law by more than $25\%$, which is equivalent to an angular momentum loss divided by two.}
\label{brakinglaw}
\end{figure}

Let us now compare these results with the semi-analytical model we developed in \citet{Reville2015b} from the parameter study of \citet{Reville2015a}. The Alfv\'en radii computed from the 3D simulations can be compared with the braking law we derived using a 2.5D axisymmetric setup in \citet{Reville2015a}:

\begin{equation}
\langle \frac{R_A}{R_{\star}} \rangle = K_3 \left( \frac{\Upsilon_{\mathrm{open}}}{(1+(f/K_4)^2)^{0.5}} \right)^m,
\end{equation}

where $K_3=0.65$, $K_4=0.06$, $m=0.3$, are fitted constants, $f=\Omega_{\star} R_{\star} / \sqrt{GM_{\star}/R_{\star}}$ is the breakup ratio and $\Upsilon_{\mathrm{open}} = \Phi_{\mathrm{open}}^2/(R_{\star}^2 \dot{M} v_{\mathrm{esc}})$ is a modified magnetization parameter \citep[see][]{UdDoula2002}. The open magnetic flux $\Phi_{\mathrm{open}}$ is computed as the unsigned magnetic flux over a spherical surface beyond the largest closed magnetic loop. The value of the open flux should be constant whatever integration surface one chooses as long as it respects this latter condition \citep[see][]{Reville2015a,Reville2015b}. Figure \ref{brakinglaw} shows how our simulations (green stars) fit this braking law. The blue line represents the braking law we derived in \citet{Reville2015a}. Our set of simulations needs a small recalibration to be modeled by our braking law. Reducing the constant $K_3$ by $15\%$ to a value of $0.55$ gives an excellent agreement for all our cases. 

The importance of the boundary conditions is illustrated in Figure \ref{brakinglaw}. The two red stars are simulations for the Sun and TYC 5164-567-1 that were made keeping fixed the initial extrapolated magnetic toroidal field inside the deepest layer of our boundary conditions. For the other (green cases), we set our self-adaptating conditions ensuring a curl free magnetic field inside the star. The deviation is around $25\%$ on the Alfv\'en radius of TYC 5164-567-1 and the Sun, which induces a large underestimation of the angular momentum loss that scales as $R_A^2$. 

In \citet{Reville2015b}, we proposed a method to compute the open flux based on an appropriate value for the source surface radius in a potential field extrapolation. We defined the optimal source surface as the source surface radius for which the potential field source surface model \citep{Schatten1969,AltschulerNewkirk1969} gives the same open flux as the simulations. To estimate this optimal source surface radius, we considered a pressure balance between the thermal and ram pressure of a spherically symmetric magneto-centrifugal wind model with the magnetic pressure of the multipolar expansion of the surface magnetic field. We demonstrated the accuracy of this estimation, using 2.5D simulations, and we show in Table \ref{table2} that it still holds for the 3D simulations performed here. The optimal ($r_{\mathrm{ss,opt}}$) and estimated ($r_{\mathrm{ss,est}}$) source surface radii are close even though $r_{\mathrm{ss,est}}$ is systematically slightly smaller.

In terms of open flux, the results from the simulation and the potential extrapolation differ by less than $10\%$ for all the cases. This good agreement is due to the right choice for the source surface radius. A potential extrapolation made with a constant $r_{\mathrm{ss}}$ would have created a large discrepancy for part of the sample, given the large variation of optimal source surface radius. The large values of $r_{\mathrm{ss,opt}}$ are consistent with the size of large coronal loops observed for the youngest stars of the sample with the largest magnetic fields that extend up to $10 R_{\star}$ (see Figure \ref{3dim}). Moreover, $r_{\mathrm{ss,opt}}$ values match the Alfv\'en points at the largest streamers' extremities and are always smaller than the effective Alfv\'en radius. This indicates there is less angular momentum at the cusp of the streamers compared to coronal holes \citep[see][]{KG2000,Garraffo2015b}.

Hence, the semi-analytical method described in \citet{Reville2015b} is likely to be successful for estimating the angular momentum loss from the mass loss and the open flux, if we adapt our formulation with the updated value of the $K_3$ constant. Going back to Figure \ref{OmegaDep}, we superimposed the mass and angular momentum loss rates given by our semi-analytical model with the dashed blue and magenta lines, respectively. The semi-analytical model overestimates the torque compared to the simulations. This can be understood simply by looking at the mass loss of the spherically symmetric wind solution used in the semi analytical model and the mass loss of the simulation (dashed and solid magenta lines). In the simulations, the mass loss is always smaller than the spherically symmetric solution, since part of the plasma is contained in closed magnetic loops that cover a large part of the stellar surface. For strong magnetic fields, it can be two to three times less than the spherically symmetric ideal case. Hence the AML is consequently smaller. It is interesting to note that the saturation does not appear in the semi-analytical model for the fastest rotators BD- 16351 and HII 296. The saturation regime we observe in the simulations complies with a linear dependency of the AML with $\Omega_{\star}$ and is thus different from the one that arises with a purely radial field in the Weber and Davis model \citep{Keppens1995}, which is used in the semi-analytical model. Hence, in addition to the intrinsic saturation of the dynamo generated magnetic field observed in X-rays, which should be contained in our magnetic maps, mass loss saturation due to confinement in large coronal loops could be involved in the saturation of angular momentum loss.

\section{Results: 3D structure of the winds}
\label{sec:vel}

\subsection{Speed distribution}

\begin{figure*}
\center
\begin{tabular}{cc}
\includegraphics[scale=0.45]{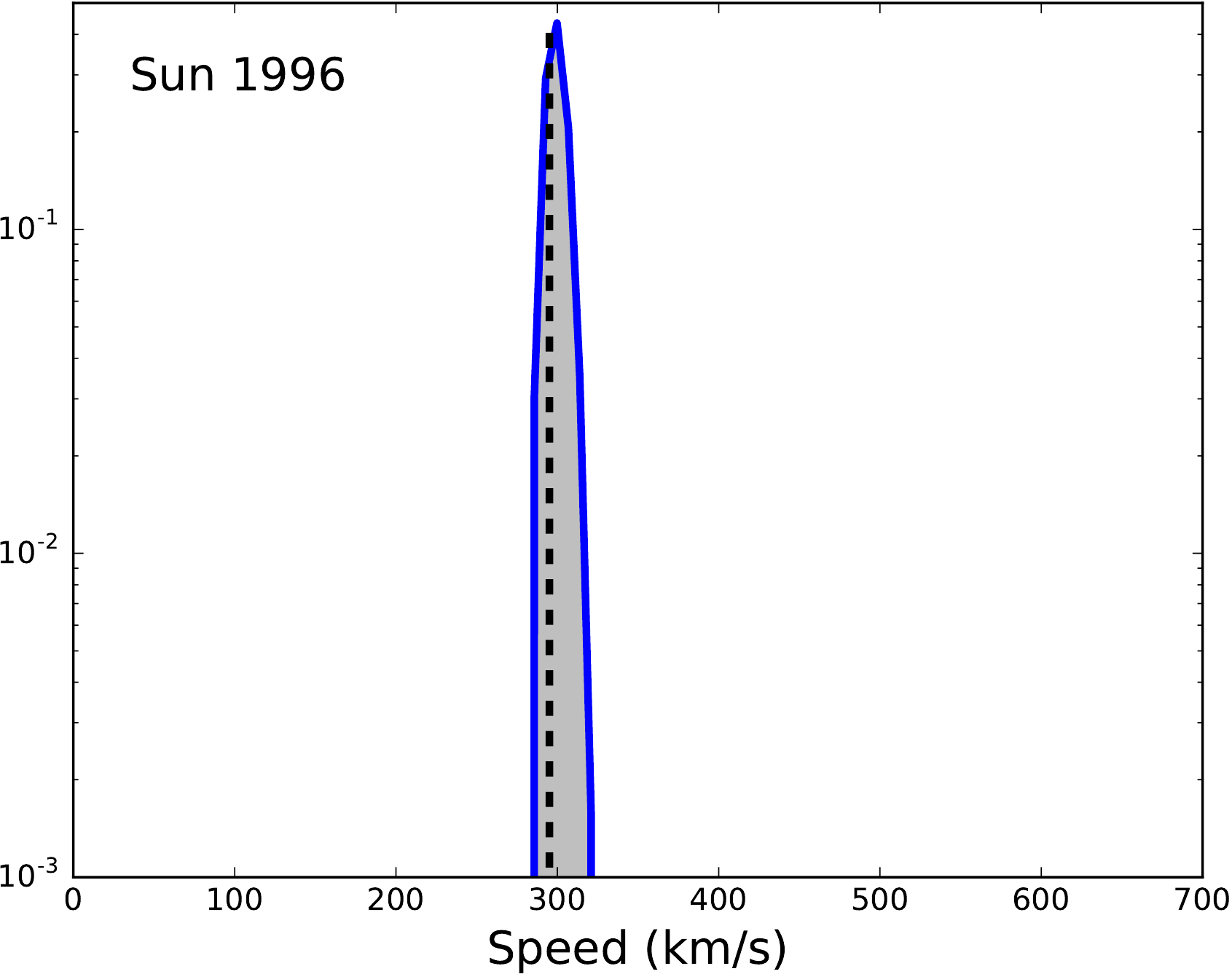} &\includegraphics[scale=0.45]{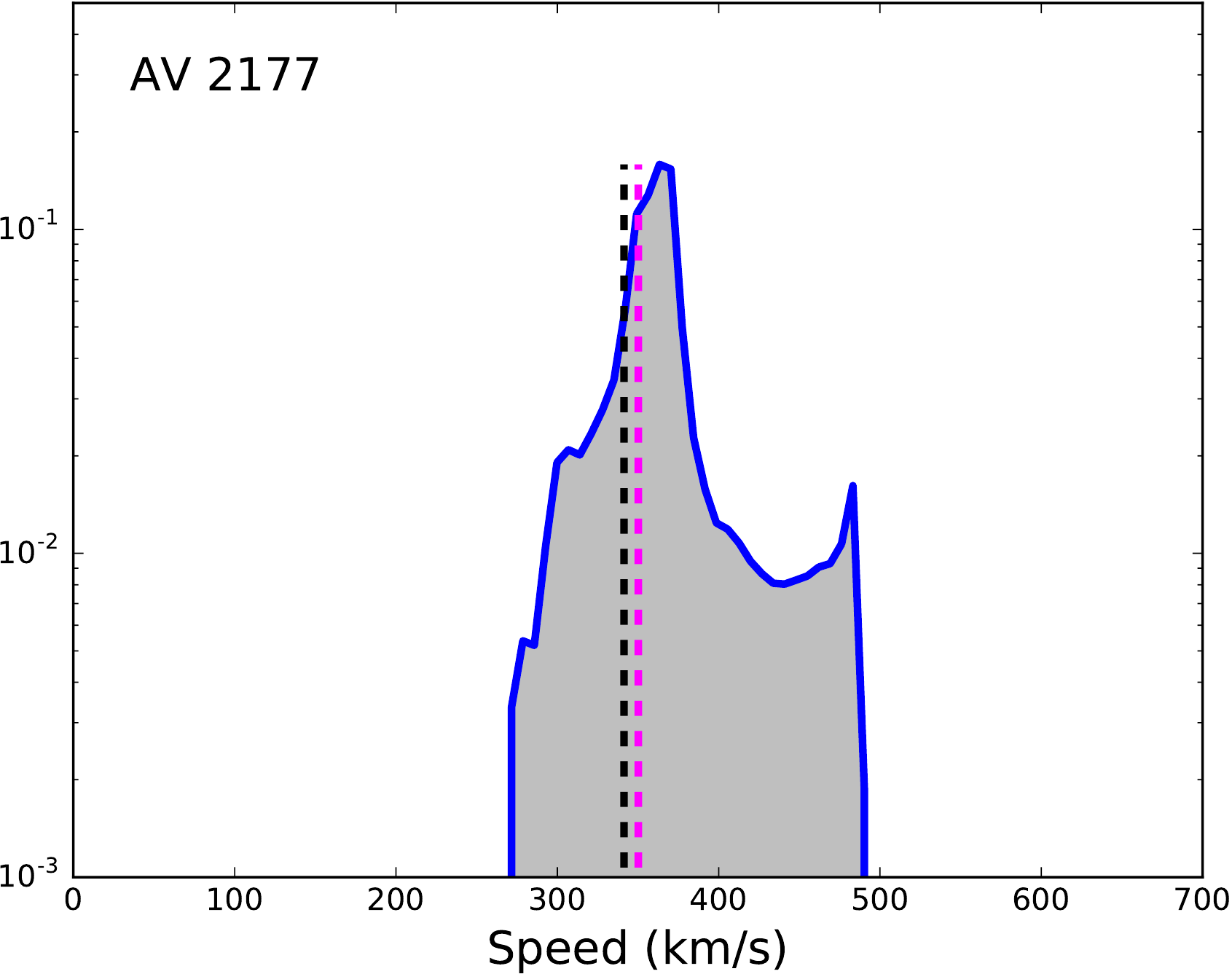} \\
\includegraphics[scale=0.45]{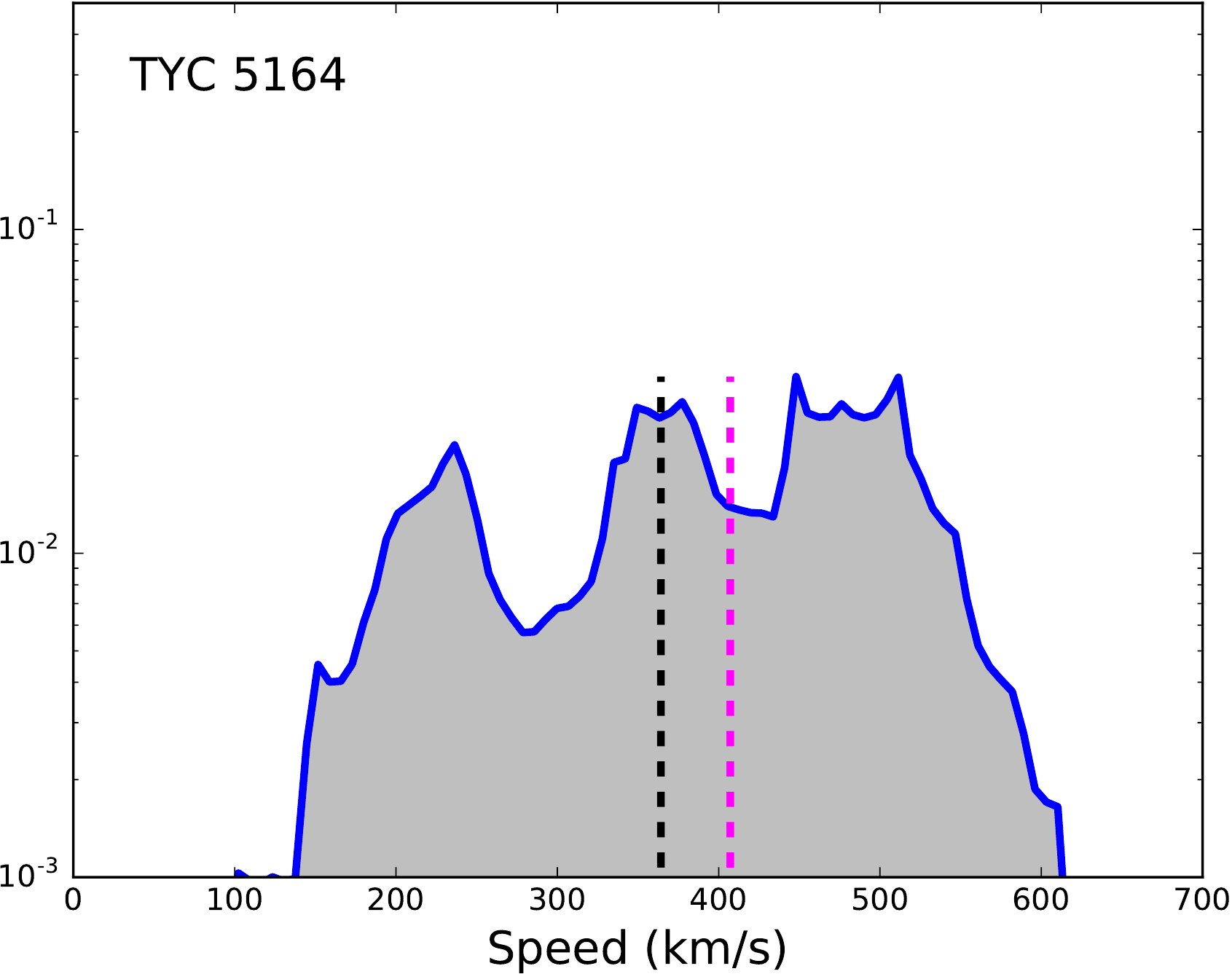} & \includegraphics[scale=0.45]{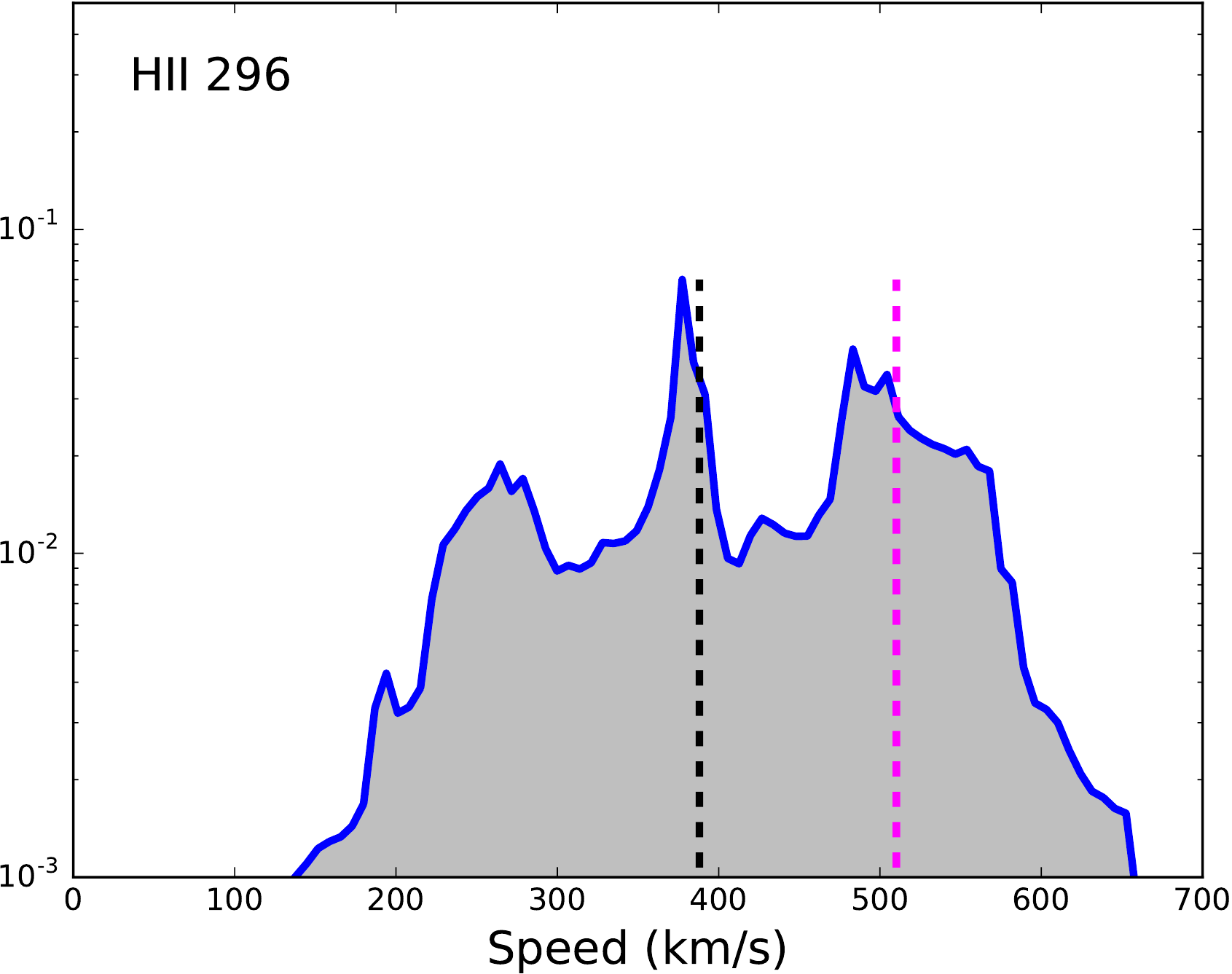} 
\end{tabular}
\caption{Speed distribution at $R=25 R_{\star}$ for 4 stars of our sample. A flattening and broadening of the distribution (all curves have a unitary integral) can be observed as the rotation and magnetic field increase. For fast rotators, three components can be identified, an average one around the polytropic solution at this radius (dashed black line), a slow component due to the slow wind emanating from streamers, and a fast component accelerated by the magneto-centrifugal effect and flux tube expansion. The speed of the magneto-centrifugal wind solution is shown in dashed magenta lines when different from the polytropic solution. The panels are sorted from left to right and top to bottom by rotation rate and not by age.}
\label{SpeedDis}
\end{figure*}

The previous section showed that global properties of young stars' winds were coherent with simpler models parametrized on 3D MHD simulations. Integrated quantities such as the mass and angular momentum loss average out the complex and 3D structure of the wind. This structure is, however, relevant when it comes to studying the interaction of those winds with other objects such as companion stars or planets \citep[see][]{Vidotto2014a,Cohen2014,Strugarek2015,doNascimento2016}

The speed distribution of all the simulations is structured with slow and fast wind components, although no fast wind related heating mechanism are included. Thermal heating is only provided by the coronal temperature and the equation of state through the choice of $\gamma=1.05$, which is scaled with the slow solar wind component. Nevertheless, if the speed distribution is narrow in our solar case, the interaction with the strong magnetic field of our fast rotators yields broader distributions and can lead to interesting dynamical properties.

Figure \ref{SpeedDis} shows the histograms of the distribution of the wind speed projected on a sphere of radius $25 R_{\star}$, for four stars of our sample. First, we observe a broadening of the speed distribution when the rotation rate of the star increases. While the speed distribution of the solar case is bracketed between $280$ km/s and $330$ km/s, for the fast rotators TYC 5164 and HII 296, we observe a flattened distribution between $200$ km/s and $650$ km/s. Since the temperature increases with rotation in our model, it is expected to get higher speeds due to the thermal driving. The theoretical speed of the polytropic solution is indicated by the black dashed line and increases with rotation. We observe, however, both higher and lower components in the stellar winds of those fast rotators.

\begin{figure*}
\center
\begin{tabular}{cc}
\textbf{Sun 1996} & \textbf{AV 2177} \\
\includegraphics[scale=0.4]{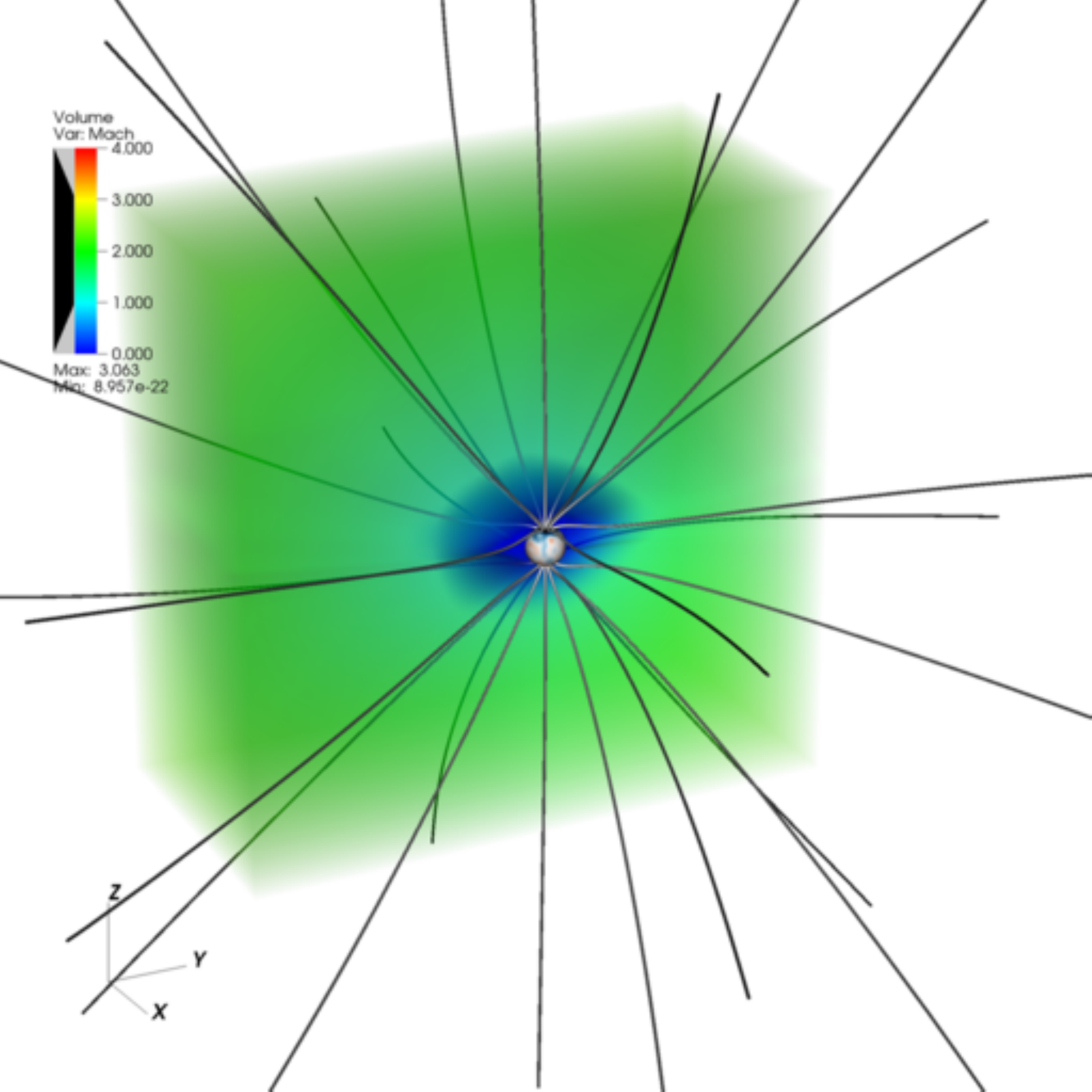} &\includegraphics[scale=0.4]{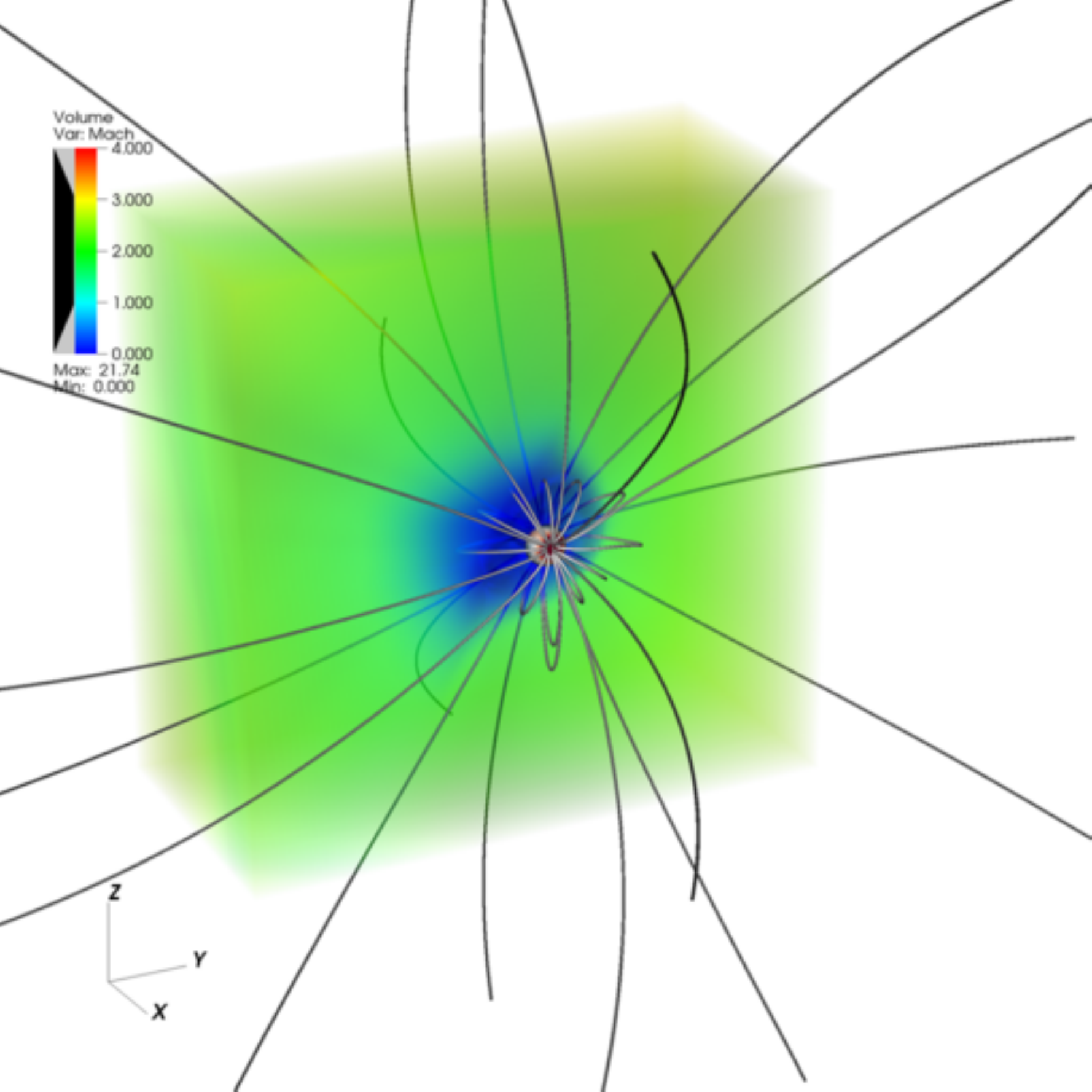} \\
\textbf{TYC 5164} & \textbf{HII 296} \\
\includegraphics[scale=0.4]{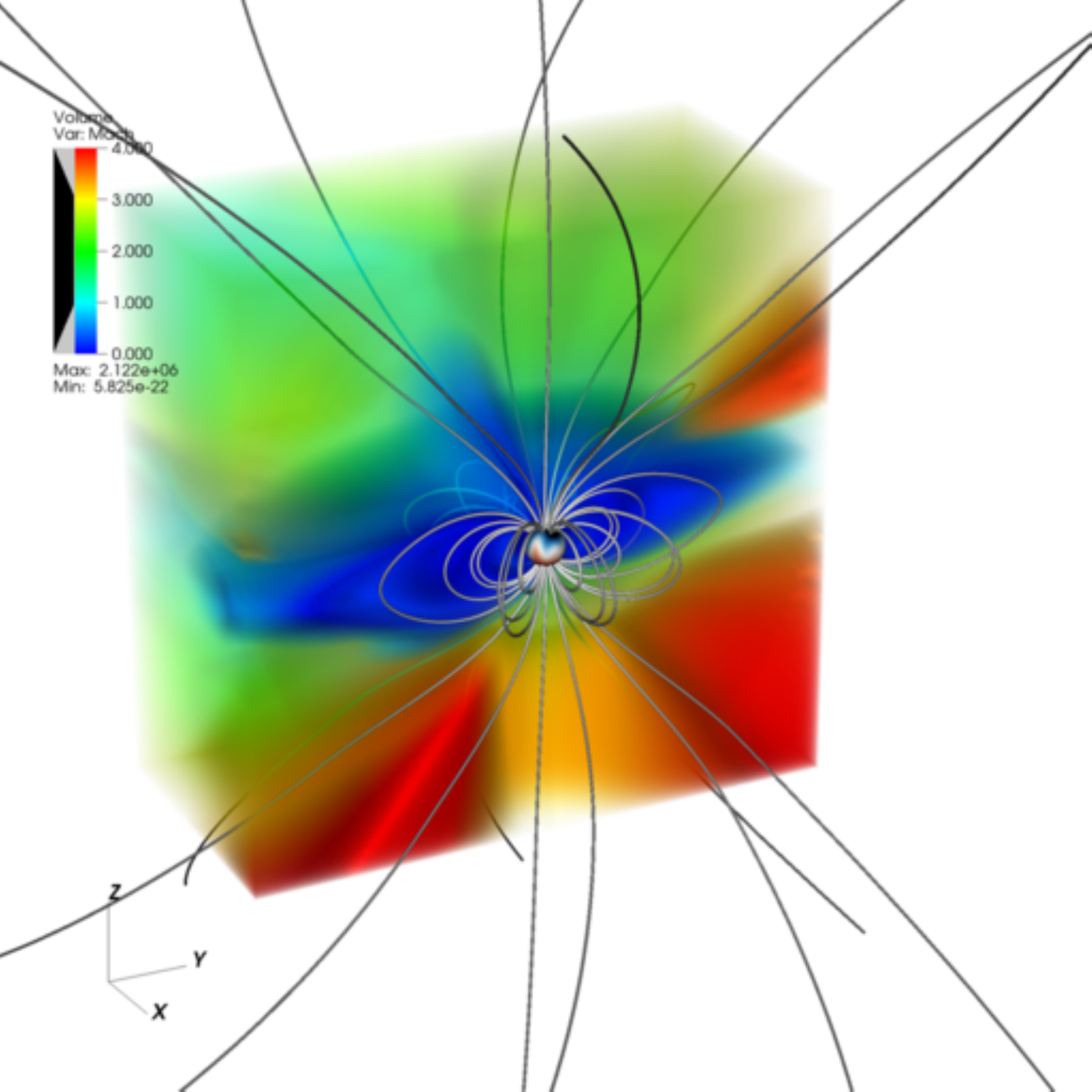} & \includegraphics[scale=0.4]{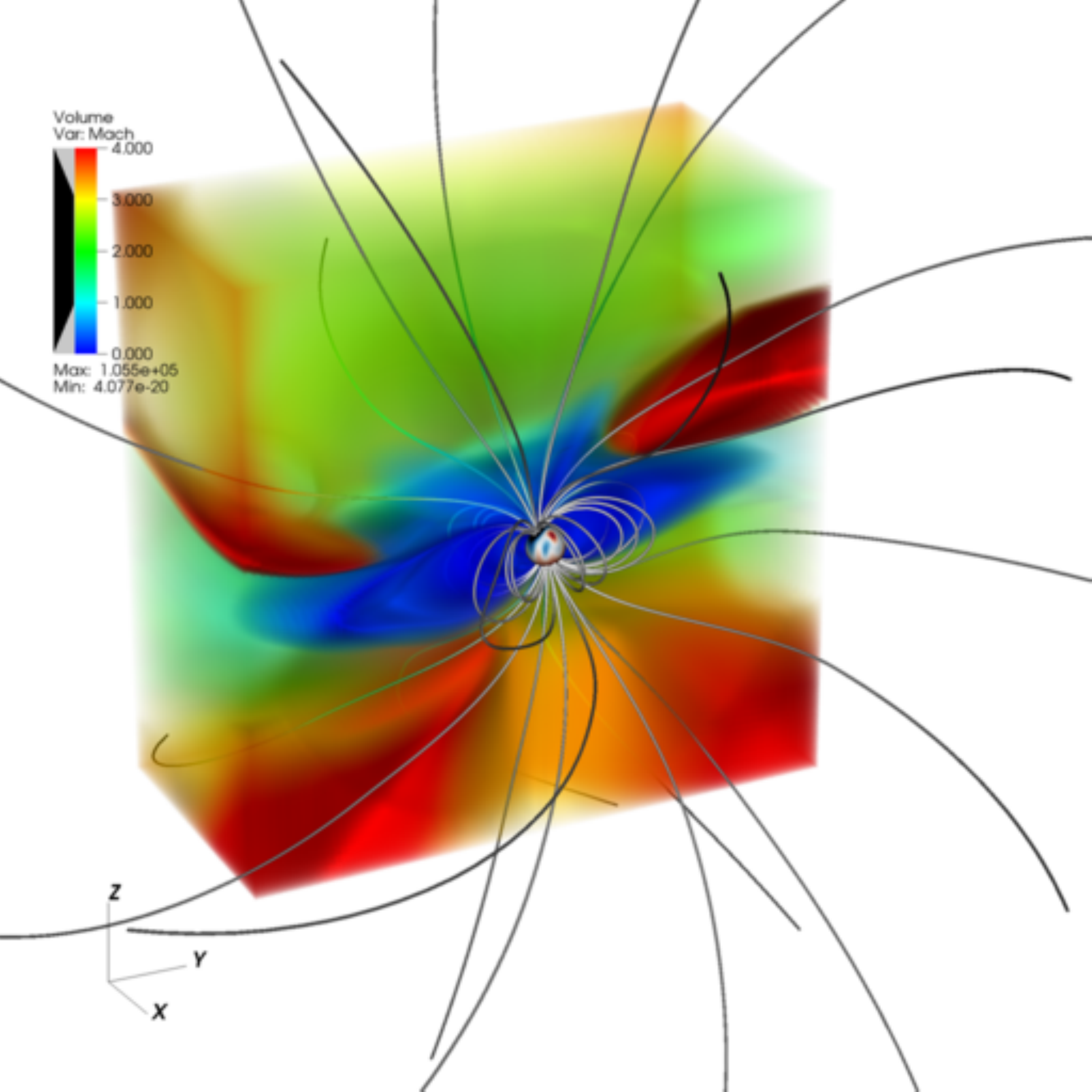} 
\end{tabular}
\caption{Volume rendering of the Mach number for the stars of Figure \ref{SpeedDis}. The three peaks in the velocity distribution of Figure \ref{SpeedDis} are visible in blue, green and red, for the slow, intermediate and fast wind component respectively.  The color scale is linear and the Mach number varies from $0$ to $4$.}
\label{MachVol}
\end{figure*}

The distribution of fast rotators is organized with three peaks. In Figure \ref{MachVol}, we show a volume rendering of the Mach number that corresponds to the four simulations of Figure \ref{SpeedDis}. We can see that, for the young stars TYC 5164-567-1 and HII 296, the trimodal distribution appears with the blue (slow component), green (intermediate component) and red (fast component) colors.  The first is due to the slow wind component emanating from the streamers. The strong dipolar magnetic fields of those stars create rings of slow winds located at the edge of the dead zones. Interestingly, those slow winds are positively correlated with the creation of currents in the simulation. The strong velocity gradient perpendicular to the dead zone boundary could be responsible for the creation of a current density even before the discontinuity in polarity that occurs beyond. This correlation needs, however, a more detailed analysis that is beyond the scope of this work.

The second peak corresponds to the theoretical speed at this distance from the star, given by the polytropic solution. Typically, this component will emanate from quiet polar regions, where the interaction with the magnetic field is the weakest. The third and fastest component appears between the streamers and slower winds blown at the poles. It seems confined in flux tubes at mid-latitudes.

The additional acceleration given to the wind is likely to come from the magneto-centrifugal effect \citep{Sakurai1985,WashShib1993,Reville2015a,Reville2015b}. Magneto-centrifugal acceleration is the consequence of the centrifugal force acting on the field lines embedded in the stellar rotation by the magnetic stress. Hence, the higher the rotation rate and the magnetic field, the higher is the magneto-centrifugal effect, which can be the dominant process in the wind acceleration \citep[see][]{Michel1969,WashShib1993}. For young stars, of periods of a few days, thermal driving and the magneto-centrifugal effect are comparable and must both be taken into account \citep{Reville2015b}. The magenta dashed line gives the wind speed obtained with the magneto-centrifugal wind solution \citep[see][for a detailed description of the solution calculation]{Sakurai1985,Reville2015b}. This value seems to be coherent for the third peak of the speed distribution of the star HII 296. Open field regions at mid-latitudes are efficiently accelerated by this process (see Figure \ref{MachVol}).

In the case of TYC 5164-567-1, the theoretical Sakurai speed is, however, lower than the observed fast peak. This could be partly explained by the fact that the magneto-centrifugal wind solution is computed with the average surface field of the star. Hot magnetic spots conveniently located at the surface have a strong enough magnetic flux to explain the lowest part of the peak. However, another effect due to fast rotation and high magnetic field is also able to accelerate a stellar wind.

\begin{figure*}
\center
\includegraphics[width=5in]{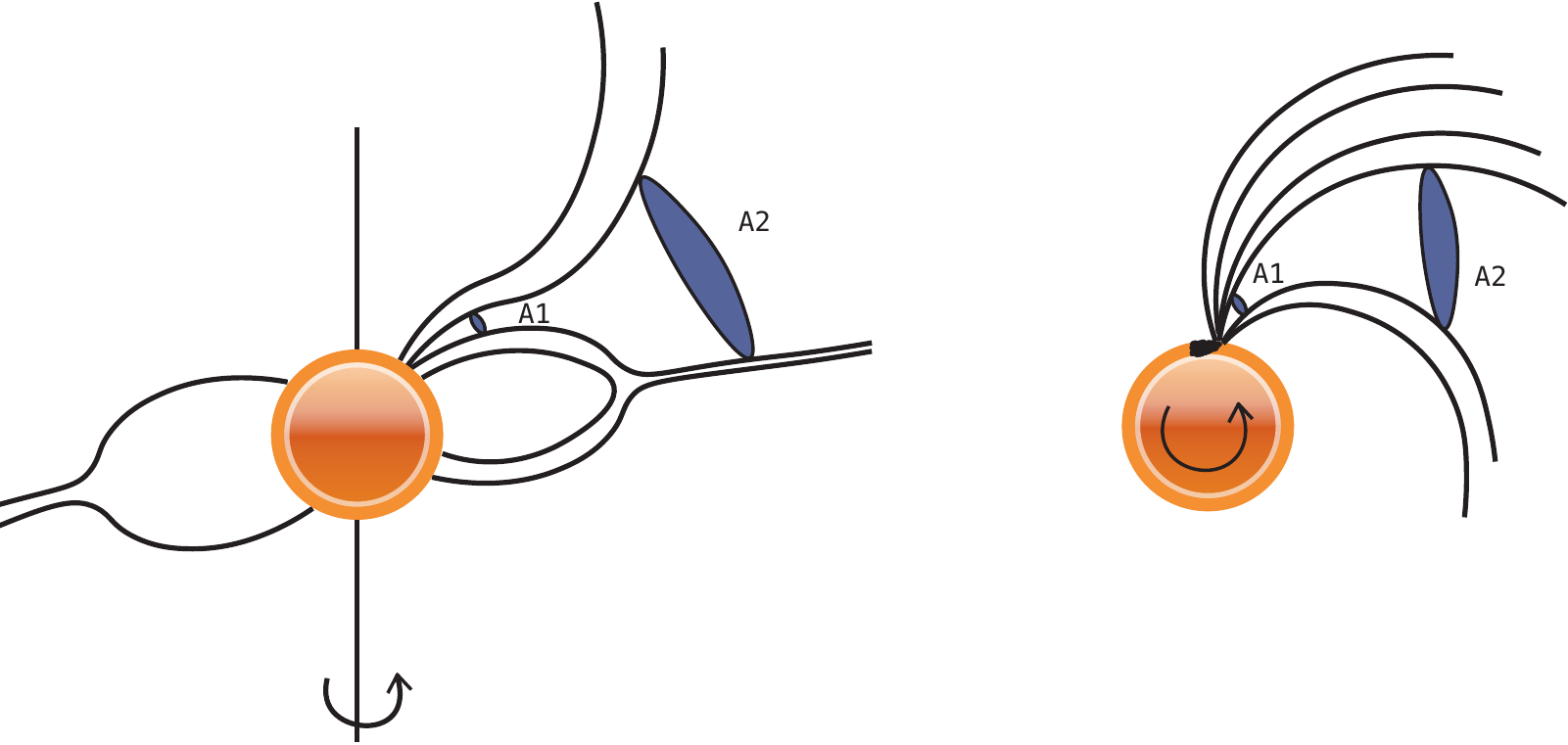}
\caption{Sketch of the latitudinal and longitudinal superradial expansion processes. On the left panel, the collimation of field lines toward the axis due to high rotation is responsible for the superradial expansion. On the right, the magnetic flux gradient at the stellar surface will generate a differential magnetic stress on both ends of the superradially expanded flux tube. See also Figure \ref{MachVol}.}
\label{ExpSketch}
\end{figure*}

For a supersonic outflow, a superradial expansion of a given magnetic flux tube will necessarily accelerate the plasma (see Appendix \ref{AppA}). Here, two effects, both due to strong magnetic fields and fast rotation, can be accounted for the expansion of flux tubes. First, a flux tube located near the pole and yet close to the streamer boundary -a typical situation with an inclined dipole topology, with a maximum configuration precisely located in longitude- will have on one side its field lines driven by the streamers and thus bent downward, and on the other side, collimation of the field lines toward the rotation axis will bend them upward. This latitudinal expansion process is illustrated in Figure \ref{ExpSketch} in the left panel.

A longitudinal expansion process can also occur when the fast flux tube originates at the boundary of the strong concentration of flux near the pole.  The magnetic stress on each side of the flux tube is different and leads to a differential efficiency of the magneto-centrifugal effect on each side. Typically, the frozen-in magnetic field line that originates inside the ``hot spot'' will rotate faster than a magnetic field lines originating outside. This effect is consequently responsible for a longitudinal expansion of the flux tube and is illustrated in the right panel of Figure \ref{ExpSketch}. 

For the fast rotators, the acceleration of fast streams clearly starts just beyond the sonic surface. We have estimated the supperradial expansion factor defined by: 

\begin{equation}
f_{\mathrm{exp}} = \left( \frac{r_c}{r} \right)^2 \frac{A}{A_c},
\end{equation}
where $r$ is the spherical radius and $A$ is the surface of the section of the magnetic flux tube, computed at $r=18 R_{\star}$ and at the sonic critical point (subscript $c$). In the case of TYC 5164-567-1, we find $f_{\mathrm{exp}}$ to increase between $6.5$ and $10$ from the outer boundary to the core of the fastest flux tube. The maximum speed observed in the core of the flux tube is around $1300$ km/s, meaning around three times the polytropic wind speed at this distance from the star. 

The broadening of the speed distribution, by the interaction of a thermally heated corona with the strong magnetic field and the fast rotation, seems to be a reliable feature in our simulations. The trimodal distribution of speeds, using polytropic and magneto-centrifugal models, could be a simple input in other models, for instance the one used to compute mass loss rates \citep[see][where they used constant stellar wind velocities in their multispecies simulations of the termination shock of astrospheres]{WoodRev2004}.

It is hard, though, to predict what could be the effect of additional accelerating mechanisms in coronal holes -that are necessary to explain the structure of the solar wind-, in the case of young stars. We can imagine that the trimodal distribution would remain, but that the separation between the slowest mode -which comes from the streamers- and the two others -which originates from coronal holes- would be widened. Also, the two fastest peaks would show higher speeds than the one obtained in our simulations.

\subsection{Slow and fast wind in the equatorial plane}

The magnetic topology of the Sun goes from a strong equatorial dipole at minimum to a more quadrupolar configuration during maximum \citep{DeRosa2012}. The heliospheric current sheet (HCS) is corrugated due to the rotationally modulated direction of the dipolar moment, and north and south sector polarities are observed at 1 AU depending on when the Earth is beneath or above it. Hence, streams of fast wind encounter slower and higher density wind that wraps the HCS. Indeed the angle of the stream $\psi = \arctan(r\Omega_{\star}/v)$ \citep{RichardsonRev2004} is a decreasing function of the velocity of the flow. Streamlines of the fast component are thus less bent and compress against slower and denser wind streams. These so-called ``Corotating Interaction Regions" (CIRs) \citep{BelcherDavis1971}, were detected in the solar system thanks to \textit{Mariner} 5. In the solar system, the spacecraft \textit{Pioneer} $10$ and $11$ have shown that $75\%$ of the CIRs have formed shocks between 3 and 5 AU \citep{SmithWolfe1976}. Those shocks dissipate the energy of the solar wind and are able, for instance, to accelerate ions \citep{RichardsonRev2004}.

\begin{figure*}
\center
\begin{tabular}{cc}
\includegraphics[scale=0.6]{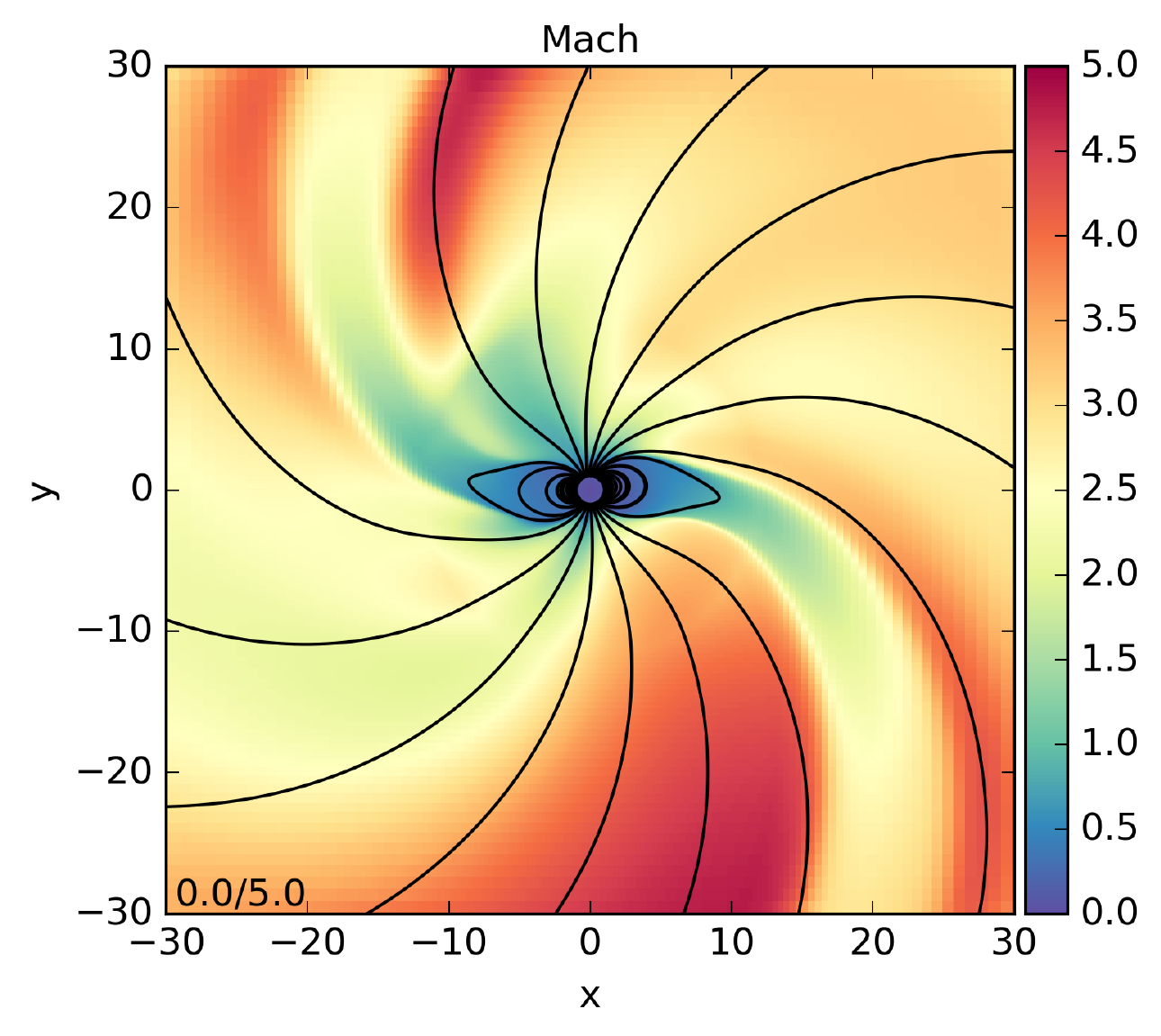} & \includegraphics[scale=0.6]{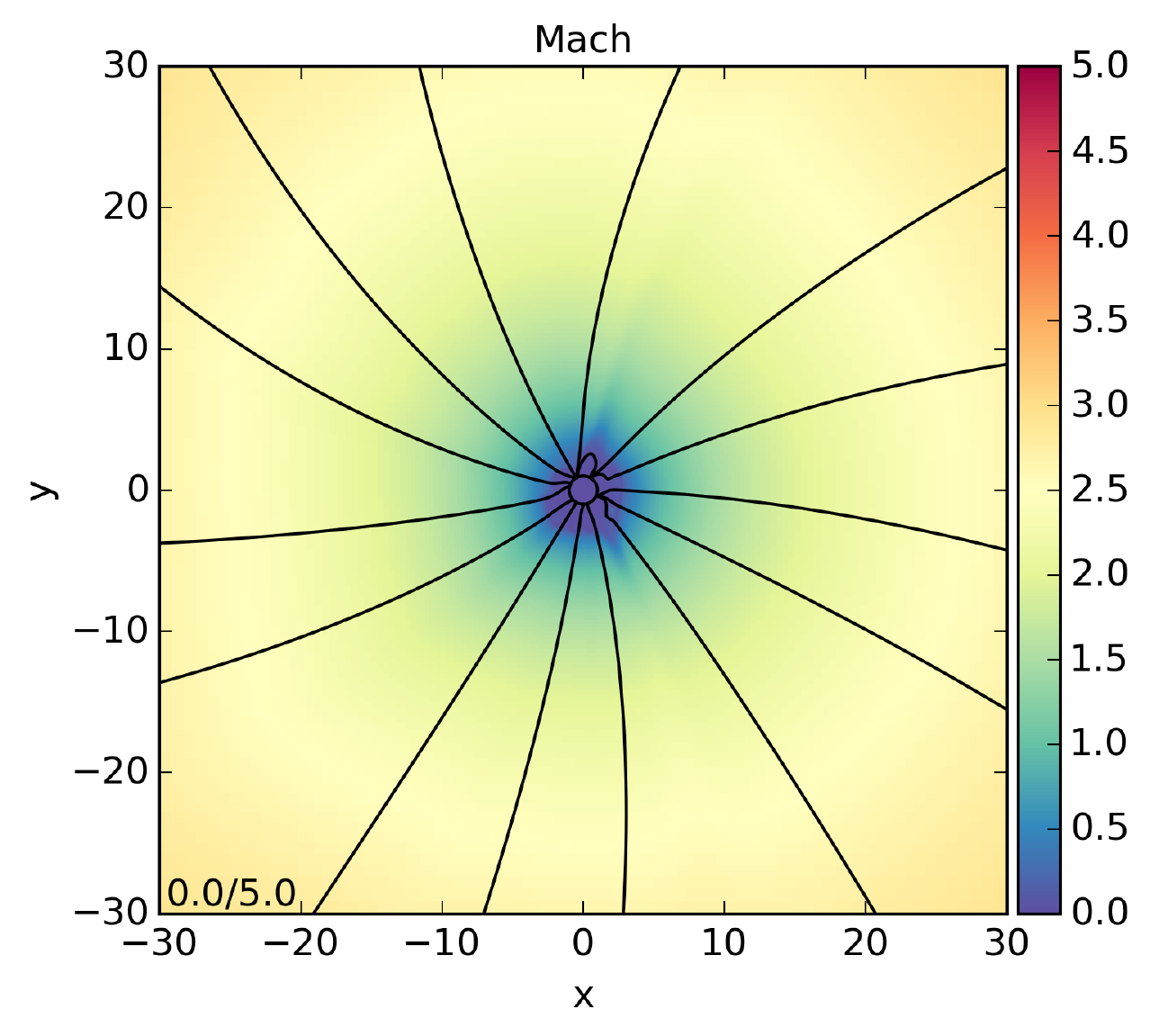}\\
\textbf{BD- 16351} & \textbf{Sun 1996}
\end{tabular}
\caption{Color maps of the Mach number and the density logarithm in the equatorial plane of BD- 16351. Magnetic field lines are plotted in black and create a Parker spiral. The fast wind component coming from coronal holes encounters the slow wind component created by streamers, and compression regions are visible at the edge of the domain.}
\label{EqPlane}
\end{figure*}

In the case of rapidly rotating stars, whose dipoles are significantly inclined  (BD- 16351 and DX Leo in our study), we should find similar features enhanced by rapid rotation and strong magnetic fields. Figure \ref{EqPlane} shows the Mach number in the equatorial plane of BD- 16351 and the Sun in 1996. The flow is structured as a Parker spiral in both cases \citep{Parker1958}. Streamers can be seen in the equatorial plane of BD- 16351 because of inclination of its dipolar magnetic field. The slow wind can be clearly identified following the streamers in that case. The flow is drastically more uniform in the case of the (quiet) Sun, and the Parker spiral shows weaker inclination of the field lines.

\begin{figure}
\center
\includegraphics[scale=0.45]{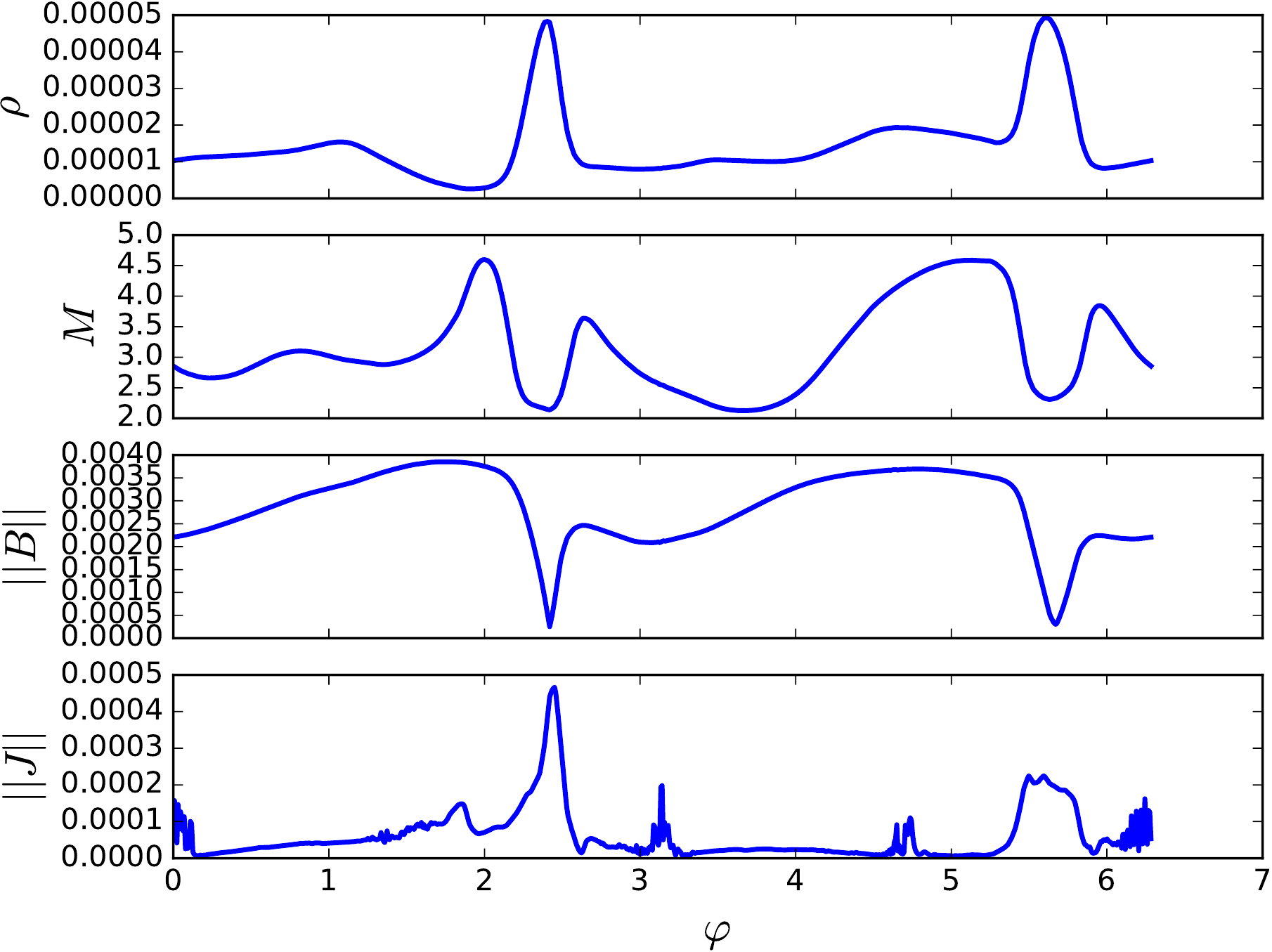}
\caption{Profiles of the density, Mach number, magnetic field and current amplitude interpolated along a $25R_{\star}$ equatorial circle in the BD- 16351 case. Peaks of density associated with troughs of velocity mark the slow wind component. Compression and acceleration occur when the fast wind encounters the slow wind component. }
\label{CIR}
\end{figure}

In the case of BD- 16351, we observe adjacent fast and slow streams that encounter each other at the edge of the domain. The direction of the field (and thus the flow) seems different for the fast stream than for the slow stream and a compression region occurs. Figure \ref{CIR} shows the profiles of the density, the Mach number, the magnetic field and the current density amplitude along a $25R_{\star}$ radius circle in the equatorial plane of BD- 16351. Two density peaks mark the slow wind components. These peaks correspond to troughs in the profiles of the Mach number and the magnetic field amplitude. The thin throat in the magnetic field profile matches the crossing of the current sheet that also corresponds to the slow wind components. This structure is similar to the heliospheric current sheet, except for its inclination which follows the strongly inclined dipole of BD- 16351.

It is interesting to look at what happens to the flow just ahead of a density peak. The magnetic field is increasing,  which means that a compression in the flow occur. However, if the density increases at first, it decreases before the peak, compensated for by an acceleration of the flow. In these regions, the wind reaches its maximum Mach number, just below 5, as shown in red in Figure \ref{EqPlane}. The velocity then drops rapidly inside the streamer, where higher (by a factor of $2$ to $5$) densities are achieved. This structure is repeated with the second streamer, beyond $\varphi = 5.5$. 

Nonetheless, the shock conditions are not yet fulfilled \citep[see, for example,][]{PlasmaPhys} and the solution would likely require us to extend it in a larger domain to produce shocks. In our simulations, the discontinuities are separated by a contact layer with no matter exchanged between them. From a numerical point of view, the HLLE solver that we use is one of the most diffusive approximate Riemann solver and may not be able to correctly capture such an oblique shock. It is likely though that shocks will form much earlier in the case of fast rotators, simply because of the enhanced helicity of the Parker spiral, thus changing the energetic budget in the astrospheres of young suns.

CIRs in the solar system are thought to be the most frequent cause for geomagnetic storms \citep{Yermolaev2012}. Our solar case does not show such features because we lack an additional heating mechanism for the acceleration of the fast wind. Yet, we can  extrapolate our results stating that younger stars are likely to create more CIRs that shock in the interplanetary medium, adding up to other dynamical events that are thought to be enhanced in the environment of active stars, such as flares and coronal mass ejections \citep{Schrijver2012}.

\section{Discussions and Perspectives}
\label{sec:ccl}

Our study addresses 3D simulations constrained by spectropolarimetric observations of magnetic fields in the context of stellar evolution, along the main sequence. Our findings can be divided into two parts. First, considering the global and integrated properties, the mass and angular momentum loss follow simple prescriptions thanks to an appropriate modeling of the evolution of the temperature and the coronal density with the rotation rate. An angular momentum proportional to the $\Omega_{\star}^3$ can be obtained using our prescriptions with observed magnetic field amplitudes, which is required to observe the convergence of spin rates on the Skumanich law.

Our 3D simulations follow the braking law we derived in \citet{Reville2015a} if the $K_3$ constant is reduced by $15\%$. This can be understood because our simulations are made with a larger (and different for each case) coronal temperature compared to this previous work. Hence, the wind described is here faster, and for a given magnetic field strength, the Alfv\'en radius is closer to the star. However, the fit shows little deviation, and one constant $K_3$ in the scaling law is able to describe the whole range of temperature of our sample. A more general braking law should quantify the influence of the temperature on $K_3$, since the variation remain limited in this study. Also, the use of a fully 3D geometry could be involved in the variation of $K_3$. Nonetheless, because this formulation expresses the dependence of the Alfv\'en radius on a magnetization parameter that includes the thermodynamics of the wind, it is likely to be valid for a wide range of magnetic fields, rotation rates, coronal temperatures and densities, if one allows a small dependence of $K_3$ on the temperature. 

With this adapted formulation, the semi-analytical we developed in \citet{Reville2015b} can be applied. The estimation method of the open flux of the simulation with a potential extrapolation that was tested on 2.5D configurations is perfectly operational with 3D non-axisymmetric fields. Our semi-analytical model is consequently able to estimate closely the evolution of angular momentum with the rotational period, as long as the mass loss rate of the simulations does not deviate too much from the spherically symmetric value used in the semi-analytical model. This deviation grows as the stars rotate faster and possess more intense magnetic fields. Large coronal loops are able to confine more plasma, and the mass loss seems to plateau for $\Omega_{\star} > 8 \Omega_{\odot}$.

This behavior has consequences on the angular momentum loss that shows signs of saturation beyond this rotation rate, whose value is coherent with the saturation value used in rotation evolution models for K-type stars \citep{GalletBouvier2015}. Although the saturation of angular momentum loss is often associated with the saturation observed in the X-ray fluxes, the precise process behind this saturation remains unknown. Some works have suggested a stochastic change of the dynamo process generating the magnetic field could be involved in a topology switch from small scales to large scales that turns on the $\Omega_{\star}^3$ braking law \citep{Barnes2003,Brown2014,Garraffo2015b}. However, no such transition is observed in our sample, as all our stars possess a strong dipolar field. In our simulations, the AML saturation seems to be due to the confinement of the outflow in large coronal loops that reduces the mass loss, which can be associated with the dependence of the wind braking on the filling factor \citep[see][]{CranmerSaar2011,GalletBouvier2013,GalletBouvier2015}. These results need, however, to be confirmed by more simulations of fast rotators and are likely to be highly dependent on the prescriptions we used for the coronal base densities and temperature. 

The mass loss rate of young stars in our study, although up to $6$ to $9$ times the solar one \footnote{In table \ref{table2}, we have considered an upper value of the solar mass loss rate $\dot{M} = 3 \times 10^{-14} M_{\odot}/$yr.}, does not reach the highest values derived in \citet{Wood2005a}. A much higher dependence on rotation, for either the temperature or the coronal density would have been necessary to observe $100 \dot{M}_{\odot}$ values in our simulations. Our semi analytical model could be used to study the influence of different prescriptions on the variation of $\dot{J}$. Change in the exponents of the evolution laws, or the solar initial values, could be tested. A more physical description of the stellar wind acceleration, driven for example by Alfv\'en waves turbulence \citep[see][]{Suzuki2006,MatsumotoSuzuki2012,Suzuki2013a,Sokolov2013,Oran2013,Lionello2014} can help to understand how coronal parameters evolve with age. Future works will be dedicated to including such processes in our simulations.

To solve this issue, more observations are also critical. In the work of \citet{Wood2005a} \citep[see also][]{WoodRev2004}, the analysis of the Ly$\alpha$ absorption spectra is coupled with numerical simulations of the terminal shock, where the wind speed is kept constant at the slow solar wind value around $400$ km/s. Our study, and this is the second part of our findings, brings more accurate constraints on the wind velocity amplitudes and distribution for solar-like stars that could be used to improve those calculations.

Indeed, several studies have now shown that the wind speeds are likely to increase for young stars, mostly because of higher coronal temperature and magneto-centrifugal acceleration \citep{WashShib1993,HolzwarthJardine2007,Matt2012,Suzuki2013b,Reville2015a,Reville2015b}. We show that the speed distribution of young and active stars follows a trimodal structure due to the interaction with the magnetic field. Moreover, the one-dimensional magneto-centrifugal wind solution \citep{WeberDavis1967,Sakurai1985} is at best a low estimate of the fastest component of the wind, as other magnetic processes are able to accelerate the wind. For instance, we observe in our simulations fast streams in the vicinity of the star caused by latitudinal or longitudinal superradial expansion of flux tubes due to the fast rotation and the non-axisymmetry of the magnetic field. 

As slow and fast winds exist around solar-like stars, they can interact in the equatorial plane, creating Corotating Interaction Regions, which could be more common for younger stars and could be forming shocks closer to the star (the usual distance observed in the solar system is between $1$ and $5$ AU). This could have consequences on the energetics of expanding stellar winds of young stars and on exoplanetary space weather. The current sheet of young stars is also strongly corrugated when the dipole is inclined and polarity variations occur closer to one another, which has important consequences on the interaction with close-in planets, especially when they are within the Alfv\'en surface \citep{Strugarek2015}. This is, however, a very preliminary study and a more focused work needs to be done in that sense, improving, for instance, the numerical scheme and shock absorbing method.

\section{acknowledgments}

We thank Sean Matt and J\'er\^{o}me Bouvier for continuous and fruitful discussions. We acknowledge funding by the ANR Blanc TOUPIES SIMI5-6 020 01, the ERC STARS2 207430, CNES via Solar Orbiter. High performance computations were performed on the French machines Turing (IDRIS) and Curie (TGCC), within the GENCI 1623 program. We are grateful to Andrea Mignone and his team at University of Torino for the development and the maintenance of the PLUTO code. 

AS is a National Postdoctoral Fellow at the Canadian Institute of Theoretical Astrophysics (CITA) and acknowledges support from the Natural Sciences and Engineering Research Council of Canada.

\begin{appendix}

\section{Flux tube expansion in the supersonic regime}
\label{AppA}

The mass conservation within a flux tube reduces to the equation

\begin{equation}
\frac{1}{v}\frac{dv}{dr}+\frac{1}{\rho}\frac{d\rho}{dr}+\frac{1}{A}\frac{dA}{dr}=0,
\end{equation}

where $A$ is the section of the flux tube. The momentum equation, if we consider an isothermal flow for simplicity, can then be written as

\begin{equation}
\left(v - \frac{c_s^2}{v}\right) \frac{dv}{dr} = \frac{c_s^2}{A} \frac{dA}{dr} - \frac{v_{\mathrm{kep}}^2 (r_c) r_c}{r^2},
\end{equation}
where $c_s$ is the constant sound speed, $v_{\mathrm{kep}}(r) =  \sqrt{GM_{\star}/r}$ is the Keplerian velocity and $r_c$ is the critical radius where the wind becomes supersonic.

In terms of Mach number $M=u/c_s$, we can write

\begin{equation}
\left(M - \frac{1}{M}\right) \frac{dM}{dr} = \frac{1}{A} \frac{dA}{dr} - \frac{v_{\mathrm{kep}}^2 (r_c) r_c}{c_s^2 r^2}.
\end{equation}

When the wind is subsonic ($M<1$), the term $(M-1/M)$ is negative, and thus what matters for the acceleration of the wind is the sign of the term on the right hand side. The expansion of the flux tube can be locally described by  $A \propto r^{\alpha}$ and yields $(1/A) dA/dr = \alpha/r$. The spherically symmetric solution, which corresponds to the case $A \propto r^2$, and $\alpha=2$ gives

\begin{equation}
\frac{2}{r} \leq \frac{v_{\mathrm{kep}}^2 (r_c) r_c}{c_s^2 r^2},
\end{equation}

as the Parker solution is always accelerating. It is shown that superradial expansion is globally anti-correlated with the wind speed \citep{WangSheeley1990}, because of the local inversion of the latter inequality below the sonic surface. Nonetheless, superradial expansion can accelerate the outflow as long as $(1/A)dA/dr$ remains smaller than $v_{\mathrm{kep}}^2 (r_c) r_c/(c_s^2 r^2)$, for instance if the superradial expansion is located near the surface or well below the sonic point \citep[see][]{Velli2010}.

However, in the supersonic regime, a superradial expansion will necessarily accelerate the outflow. Indeed, as $\alpha \geq 2$, and since the right hand side is positive in the radial case, any superradial expansion will grow this term, and $dM/dr$ will be larger. 

The treatment of this problem without the isothermal approximation is more complex \citep[see][]{KoppHolzer1976}. In our study, the sound speed variation may not be negligible, but the qualitative behavior remains and a quantitative analysis is left for future works.

\end{appendix}

\end{document}